\documentclass[11pt,onecolumn,superscriptaddress]{revtex4}

\usepackage[dvips]{graphicx}
\usepackage{amssymb,amsfonts,amsmath}
\usepackage{color}

\newcommand{\um}{\textrm{$\mu$m}}
\newcommand{\nm}{\textrm{nm}} 
\newcommand{\scc}{\textrm{s}}
\newcommand{\nM}{\textrm{nM}}
\newcommand{\M}{\textrm{M}}
\newcommand{\kcm}{\textrm{kcal/mol}}
\newcommand{\nsb}{\textrm{NSB}}
\newcommand{\lp}{\textrm{loop}}
\newcommand{\ulp}{\textrm{unloop}}
\newcommand{\tet}{\textrm{tet}}

\newcommand{\ci}{\textrm{CI}}
\newcommand{\cro}{\textrm{Cro}}
\newcommand{\lys}{\textrm{lys}}
\newcommand{\ai}{\textrm{ai}}
\newcommand{\dd}{\textrm{D}}

\begin{document}

\title{DNA looping provides stability and robustness to the bacteriophage $\lambda$ switch}

\author{Marco J. Morelli}
\affiliation{FOM Institute for Atomic and Molecular Physics, Kruislaan 407, 1098 SJ Amsterdam, The Netherlands}
\affiliation{Current address: Division of Ecology and Evolutionary Biology, University of Glasgow, Glasgow G12 8QQ, UK},

\author{Pieter Rein ten Wolde}
\affiliation{FOM Institute for Atomic and Molecular Physics, Kruislaan 407, 1098 SJ Amsterdam, The Netherlands}

\author{Rosalind J. Allen}
\affiliation{SUPA, School of Physics and Astronomy, The University of Edinburgh, James Clerk Maxwell Building, The King's Buildings, Mayfield Road, Edinburgh EH9 3JZ, UK}

\begin{abstract}
  The bistable gene regulatory switch controlling the transition from lysogeny
  to lysis in bacteriophage $\lambda$ presents a unique challenge to
  quantitative modeling. Despite extensive
  characterization of this regulatory network, the origin of the extreme
  stability of the lysogenic state remains unclear. We have
  constructed a stochastic model for this switch. Using Forward Flux
  Sampling simulations, we show that this model predicts an extremely low rate of
  spontaneous prophage induction in a {\em{recA}} mutant, in agreement
  with experimental observations. In our model, the DNA loop formed by
  octamerization of CI bound to the $O_L$ and $O_R$ operator regions
  is crucial for stability, allowing the lysogenic state to remain
  stable even when a large fraction of the total CI is depleted by
  nonspecific binding to genomic DNA. DNA looping also ensures that
  the switch is robust to mutations in the order of the $O_R$
  binding sites. Our results suggest that DNA looping can provide a
  mechanism to maintain a stable lysogenic state in the face of a
  range of challenges including noisy gene expression, nonspecific DNA
  binding and operator site mutations.
\end{abstract}

\keywords{bacteriophage lambda | stochastic modeling | biochemical networks | genetic switch | rare events}

\maketitle

\begin{figure}[!h]
\includegraphics[width=.55\textwidth,clip=true]{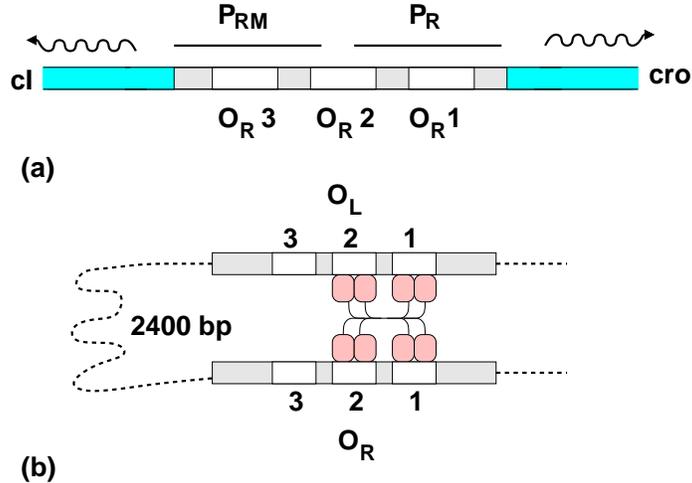}
\caption{Schematic illustration of (a): the $O_R$ region of the phage $\lambda$ switch, and (b): the DNA loop formed by octamerization of CI dimers bound of $O_R$ and $O_L$.\label{fig:diag}}
\end{figure}

The bistable developmental switch controlling the transition from
lysogeny to lysis in bacteriophage $\lambda$ is one of the best
characterized gene regulatory networks \cite{lphagebook}. In the
lysogenic state, the phage $\lambda$ genome is integrated into the
chromosome of the {\em{Escherichia coli}} host cell and is essentially dormant, due to expression of
the $cI$ repressor gene, the product of which represses $cro$ and other genes (Fig. \ref{fig:diag}). A
transition to the lytic switch state can occur in response to DNA
damage (via UV irradiation), when CI molecules are cleaved by RecA. Transcription of the $cro$ gene from $P_R$ then triggers a cascade of gene activation,
leading to phage excision, replication and cell lysis. In mutants
where this cascade is blocked, a state with elevated Cro levels is
stable for several cell generations. This is known as the anti-immune
state \cite{Calef}. A simple and intuitive explanation has been presented for this
bistability \cite{lphagebook}: in the lysogenic state, CI is
dominant and $cro$ is repressed; yet, once $cro$ begins to be
expressed, Cro dimers repress  transcription of $cI$, making the transition to lysis
inevitable. However, quantitative measurements have revealed a
puzzle: the lysogenic state of the phage $\lambda$ switch is both 
extremely stable \cite{Little99,Aurell02,Rozanov98} and robust to rewiring
of its transcriptional regulatory interactions
\cite{Little99}. These features have not yet been explained by mathematical models. Here, we present dynamical simulations of a stochastic mathematical model that reproduces this seemingly mysterious behavior. Our simulations provide evidence that DNA looping plays a key role in ensuring the stability and robustness of the switch.

The molecular interactions controlling the transcription of $cI$ and
$cro$ have been studied in great detail. CI and Cro bind as dimers to
the operator sites $O_R1$, $O_R2$ and $O_R3$ (Figure \ref{fig:diag}), which control
expression of the  {\em{cI}} and
{\em{cro}} genes from the $P_{RM}$ and $P_R$ promoters. Transcription from the unactivated $P_{RM}$ promoter is about 10 times
weaker than from $P_R$;  however, when a
CI dimer is bound at $O_R2$, transcription from $P_{RM}$ is enhanced to about the same level as from $P_R$ ({\em{i.e.}} the two promoters can compete with one another only when CI is bound at $O_R2$).  CI dimers bind preferentially and
co-operatively to the $O_R1$ and $O_R2$ sites, which overlap the $P_R$
promoter. When CI is bound to these two sites, $cro$ ($P_R$) is repressed and
$cI$ ($P_{RM}$) is activated. Cro dimers bind preferentially to $O_R3$, which
overlaps $P_{RM}$, so that when this site is occupied, $cI$ is
repressed \cite{lphagebook}. An important additional component of the network
architecture is a DNA loop which can form between the $O_R$ site and
the left operator site $O_L$, located 2400bp from $O_R$, which also
has three adjacent binding sites for CI dimers ($O_L1$, $O_L2$ and
$O_L3$). The loop is mediated by octamerization between pairs of CI
dimers bound at $O_R$ and $O_L$ \cite{Revet99}. The role of this DNA
looping interaction in the function of the phage $\lambda$ switch
remains a subject of debate
\cite{Dodd01,Dodd04,Svenningsen05,roma,anderson2008}.

Quantitative measurements have revealed several intriguing features of
the phage $\lambda$ switch. Firstly, the lysogenic state is extremely
stable \cite{Little99,Aurell02,Rozanov98}, despite the stochastic
nature of the underlying gene regulatory network
 Stochastic fluctuations in gene regulation (``noise'') might be expected to cause spontaneous transitions from the lysogenic to lytic states, even in lysogens lacking RecA; yet the
rate of these transitions is so low as to be almost unmeasurable.
Secondly, recent measurements of $P_{RM}$ and $P_R$ promoter activity
suggest that only a small fraction of the total CI in the cell is
available for binding to $O_R$
\cite{Dodd01,Dodd04,Reinitz90,Pakula86,Bakk04_1}, while other
measurements show that the total concentration of CI in the lysogen varies
dramatically from cell to cell \cite{Baek03}. Taken together, these
results suggest that the stability of the lysogen is rather
insensitive to the number of free intracellular CI molecules. Despite
this stability, transition to lysis occurs readily in wild-type phage
on UV irradiation, which leads to cleavage of CI by RecA. Finally,
and remarkably, switch function is robust to changes in the gene
network architecture itself: when the order of the three $O_R$ binding
sites is altered so that $O_R1$ is replaced by $O_R3$ or {\em{vice
    versa}}, the network remains functional \cite{Little99}.

Computer simulations should be an excellent tool for explaining this
behavior. Although stochastic simulations have successfully been used to model the initial developmental choice between lysogeny and lysis for lambda-infected cells \cite{arkin1998}, modelling spontaneous switching of an already established lysogen has proved problematic. Despite the fact that a wealth of biochemical
parameters are available for this network, no model has reproduced its
extraordinary stability and functional robustness
 \cite{Aurell02,Reinitz90,AS,Tian04,Lou07,Zhu04,Santillan04}. 
 In this paper, we present a model that takes into account the stochastic character of the chemical reactions and includes DNA looping and depletion of free CI and Cro by nonspecific binding to genomic DNA. Our model also explicitly describes the detailed dynamics of the binding of transcription factors to the promoters. Accurate computation of spontaneous switching rates for this large reaction set is achieved using the Forward Flux Sampling (FFS) rare event simulation method~\cite{Allen05,Allen06_1}, 
in combination with temporal coarse-graining of dimerization and nonspecific DNA binding reactions. Our simulations show that this stochastic model can reproduce the bistability of the switch and its robustness to operator site mutations, as well as the extreme stability of the lysogenic state, even in the presence of nonspecific DNA binding.

In this work, we study the effect on switch function of two key
parameters: the strengths of the DNA looping and nonspecific binding
interactions. We find that the DNA looping interaction plays a crucial
role. In the absence of the looping interaction, a highly stable 
lysogenic state can be achieved, but this state is very sensitive to depletion of free CI and to operator site mutations. When looping is included in the model, the lysogen is insensitive to CI depletion and robust to rearrangement of the operator sites. We conclude that DNA looping may
play an important role in allowing the phage $\lambda$ switch to
function reliably even under highly destabilizing conditions in the
host cell.

\section{The model}

Our model consists of a set of chemical
reactions, simulated using the Gillespie algorithm
 \cite{Gillespie76}. The components of the model are:
dimerization of CI and Cro proteins, binding of CI and Cro dimers to
specific DNA binding sites $O_R1$, $O_R2$, $O_R3$, $O_L1$, $O_L2$ and
$O_L3$, binding of RNA polymerase (RNAp) to promoters $P_{RM}$, $P_R$
and $P_L$, transcription of {\em{cI}} and {\em{cro}}, translation of
the corresponding mRNA transcripts, degradation of mRNA transcripts
and removal of CI and Cro monomers and dimers from the cell. Our model
also includes formation of a DNA loop between $O_R$ and $O_L$,
mediated by a CI octamer, and nonspecific binding of CI and Cro dimers
to genomic DNA. The key parameters that we vary are the strength of
the nonspecific DNA binding interaction $\Delta G_{\nsb}$ and the
strength of the DNA looping interaction $\Delta G_{\lp}$. Other
parameters are fixed using biochemical data as far as
possible. The model parameters are discussed briefly here and
described in full in the {\em Supporting Information}.
\subsection{Host cell parameters}
We assume that the {\em{E. coli}} host cell is growing rapidly (doubling time 34 min \cite{Little99}), and has 3 copies of the $O_R$ and $O_L$ operators \cite{Aurell02}, in a cell volume of 2$\um^3$ \cite{Aurell02}. The concentration of free RNAp in the cytoplasm is taken to be  50$\nM$ \cite{mcclure1983_proc}, but our conclusions are not sensitive to this parameter, as we demonstrate in the {\em Supporting Information}.
\subsection{Operator binding dynamics} Equilibrium constants from the
literature were used for CI \cite{Koblan92,Burz94_1} and Cro
\cite{Darling00} binding to $O_R1$, $O_R2$, $O_R3$, for RNAp binding
to $P_{RM}$ and $P_R$ \cite{Shea85} and for CI binding to the $O_L$
sites \cite{Dodd04}. Cro is assumed to bind to $O_L$ and $O_R$ sites
identically. The total number of possible (unlooped) configurations of the $O_R$
and $O_L$ operators are, respectively,  40 and 36. Since we are
performing dynamical simulations, we require rate constants $k_{a}$
and $k_{d}$ for association and dissociation. For all association
rates, we used the diffusion-limited value $k_{a}=4\pi D \sigma =
0.314 \um^3\scc^{-1}$ (taking the diffusion constant $D=5
\um^2\scc^{-1}$ and the molecular size $\sigma=5\nm$). The rate
constant for dissociation, in $\scc^{-1}$, was then deduced from the
equilibrium constant, using $k_{a}/k_{d}=(\exp{[-\Delta G/RT]})/(6.023
\times 10^8)\um^3$, where $k_{a}$ is in $\um^3\scc^{-1}$, $\Delta G$
is in $\kcm$, $RT=0.616\kcm$ at 37C, and $6.023 \times 10^8\um^3$ is a
volume conversion factor.
\subsection{Protein and mRNA production and removal} We model transcription as a single reaction in which an mRNA molecule is produced when RNAp is bound to a promoter. $P_{RM}$ activity is enhanced when a CI dimer is bound at $O_R2$. Transcription rates are 0.014$\scc^{-1}$ for $P_R$, 0.001$\scc^{-1}$ for unstimulated $P_{RM}$ and 0.011$\scc^{-1}$ for stimulated $P_{RM}$ \cite{Shea85,McClure82,Hwang88,Li97}. All mRNA transcripts are degraded with a half-life of 2 mins. Translation and protein folding are combined into a single step. The model produces a statistical distribution for the number of proteins produced per transcript, which is governed by the balance betwen the translation and mRNA degradation rates. The average of this distribution (the ``burst size'') is 6 and 20 for CI and Cro respectively. The burst size for Cro follows ref. \cite{Aurell02}. The value for CI is based on the observation that the CI ribosome binding site (RBS) is $\sim 7$-fold weaker than that of LacZ \cite{dodd_pc}, and that the LacZ burst size is 30-40 \cite{kennell,sorensen}. A somewhat weaker CI RBS was observed by Shean and Gottesman \cite{Shean92}.  Protein monomers and dimers are removed  with rate constant $\ln (2) / T$, where the cell cycle time $T=$34min \cite{Little99}. We also include active degradation of Cro monomers with half-life 42min \cite{Pakula86,Aurell02}.
\subsection{Dimerization} Dimerization free energies are taken to be $-11\kcm$ for CI \cite{Burz94,Beckett91} and $-8.7\kcm$ for Cro \cite{Darling00_1}. The association reaction is again assumed to be diffusion-limited.  To increase the efficiency of our simulations, we coarse-grain the monomer-monomer association and  dissociation reactions for both CI and Cro \cite{morelli2008}, as described in   the {\em{Supporting Information}}. 
\subsection{DNA looping} When an  $O_R$ and an $O_L$ operator each carry at
least two adjacent CI dimers (at the 1-2 or 2-3 sites), these
operators can associate to form a
``looped state'', with rate $k_\lp$, which dissociates with rate
$k_\ulp$. The total number of possible looped states is 49. Binding of CI dimers to the two non-octamerized sites in a loop occurs with a cooperativity factor $\exp{[-\Delta G_{\tet}/RT]}$ where
$\Delta G_{\tet}=-3 \kcm$ \cite{Dodd04}. Because we assume fast growth of the host cell, our model contains 3 copies of the host genome and 3 copies of the phage $\lambda$ switch. Since the loop is much longer than the persistence length of DNA, we assume that any $O_R-O_L$ combination  can form a loop. The strength and dynamics of the DNA loop {\em{in vivo}} are
 unknown. We therefore test the effects of DNA looping on the network
 behavior, by varying the ratio $k_\lp/k_\ulp \equiv \exp{[-\Delta
   G_{\lp}/RT]}$. We generally assume a fixed value $62.1\scc^{-1}$
 for $k_\lp$ (arising from considerations of polymer dynamics as discussed in the {\em Supporting Information}), but we
 find that only the ratio is important.  
\subsection{Nonspecific DNA binding dynamics} We model nonspecific DNA
binding \cite{Dodd04,Bakk04_1} by including in our reaction set
association and dissociation of CI and Cro dimers to $10^7$ genomic
DNA sites, corresponding to 2-3 copies of the bacterial genome. The
association rate $k_a$ is assumed to be diffusion-limited, and we assume identical nonspecific binding affinities for CI and Cro. Nonspecifically bound dimers are removed from the cell with rate constant $\ln(2)/T$. We do not
model nonspecific binding of RNAp, since our value of 50$\nM$ corresponds to the {\em{free}} RNAp concentration \cite{mcclure1983_proc}. To investigate the effects of
nonspecific DNA binding on the model switch, we vary the parameter
$\Delta G_{\nsb}$ where $k_{a}/k_{d}=(\exp{[-\Delta
  G_\nsb/RT]})/(6.023 \times 10^8)\um^3$. We assume that these
reactions are fast compared to the other reactions in the network, and
can be coarse-grained \cite{morelli2008} as described in 
the {\em Supporting Information}.

\section{Bistability}

\begin{figure}[!h]
\includegraphics[width=.55\textwidth,clip=true]{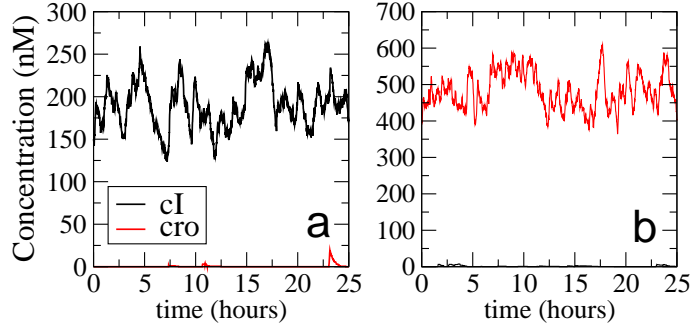}
\caption{\label{fig:traces} The model switch is bistable. Typical simulation trajectories are shown for $\Delta G_{\lp}=-3.7\kcm$ and $\Delta G_{\nsb}=-4.1\kcm$. (a): A simulation initiated with 150$\nM$ CI and no Cro molecules remains in the CI-rich state for many hours (b): A simulation initiated with 400$\nM$ Cro and no CI molecules remains in the Cro-rich state.}
\end{figure}

\begin{figure}[!h]
\includegraphics[width=.4\textwidth,clip=true]{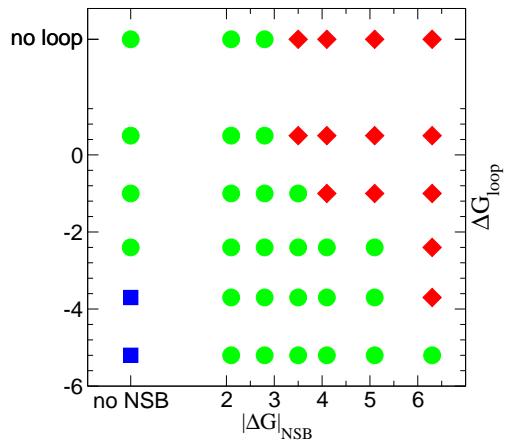}
\caption{\label{fig:phdiag} Bistability of the model switch as a
  function of the nonspecific binding strength $\Delta G_{\nsb}$ and
  the DNA looping strength $\Delta G_{\lp}$. Blue squares represent
  parameter sets for which only the lysogenic state is stable (no
  stable anti-immune state); for red diamonds only the anti-immune state
  is stable (no stable lysogen), while green circles represent parameter
  combinations where the model switch is bistable.}
\end{figure}

Our model represents a mutant phage in which the lytic pathway is
nonfunctional, since we do not model the lytic genes downstream of
{\em{cro}} \cite{Calef}. Such mutants show two stable states: the very stable
lysogenic state, with high CI and low Cro levels, and the anti-immune
state, with elevated Cro and little CI, which is less stable, but is
nevertheless maintained for several generations \cite{Calef}. For our
model to reproduce this bistability, simulations initiated in either
of the lysogenic or anti-immune states should remain stable, only rarely making a spontaneous transition to the
other state. Figure \ref{fig:traces} shows that this is indeed the
case: in Figure \ref{fig:traces}a, a simulation initiated in the
lysogenic state remains in that state, while in Figure
\ref{fig:traces}b, a simulation run with the same parameters but
initiated with a high Cro concentration and little CI remains in the
anti-immune state. Figure \ref{fig:phdiag} shows the range of
parameter values for which our model shows bistability. Our
simulations give steady state concentrations of CI and Cro in good
agreement with measured values for the lysogenic and anti-immune
states. For the lysogenic state, we obtain $\sim 200-400$ CI per cell,
compared to a measured value of 220 \cite{Reichardt71}. For the
anti-immune state, Cro per cell ranges from $\sim 250 -900$ (depending on $\Delta G_{\nsb}$), corresponding to concentrations of $200-750\nM$, compared to a measured concentration of $\sim 400\nM$ \cite{Reinitz90}. These
values are presented as functions of $\Delta
G_{\lp}$ and $\Delta G_{\nsb}$ in the {\em{Supporting Information}}. In
our model, the DNA looping interaction decreases the lysogenic CI
concentration by as much as a factor of 2, in agreement with the observations of Dodd {\em{et al}}
\cite{Dodd01,Dodd04}. We have also simulated mutants without the
$O_R3$ or the $O_L3$ binding sites, corresponding approximately to
the OR3-r1 and OL3-4 mutants of refs \cite{Dodd01},
\cite{Dodd04} and \cite{anderson2008}. For these mutants, we find lysogenic CI
concentrations 1.9 and 1.8 times the wild-type values
(for $\Delta G_\lp=-3.7\kcm$ and $\Delta G_\nsb=-2.8\kcm$), in
reasonable agreement with the results of ref \cite{Dodd04} (factors of 2.8 and 3 respectively), even though our model does not include the
up-regulation of $P_{RM}$ by the loop  identified in ref
\cite{anderson2008}, or the effect on $P_{RM}$ of the OR3-r1 mutation. For the same parameters, the $P_R$ promoter is repressed by a factor of $1.6$ in the anti-immune state, in good agreement with ref. \cite{Svenningsen05}.

\section{Extreme stability of the lysogenic state}

The spontaneous switching rate from lysogeny to lysis in {\em{recA--}} mutants is so low as to be almost undetectable, being less than $2 \times 10^{-9}$ per cell per generation \cite{Little99,Aurell02}. Reproducing such low spontaneous switching rates while maintaining bistability provides an extreme challenge for a computational model \cite{Aurell02}. We quantified the stability of the lysogenic and anti-immune states for our model, using the forward flux sampling (FFS) rare event simulation method \cite{Allen05,Allen06_1}. This method allows us to compute the rate of spontaneous switch flips, even though such flips would hardly ever be observed in a typical ``brute-force'' simulation run. FFS uses a  series of interfaces (defined by an order parameter) between the initial (CI-rich) and final (Cro-rich) states to split a switching event into a number of more likely transitions between successive interfaces. Here, our order parameter was  the difference between the total number of Cro and CI molecules. This method has previously been used for a simple model of a mutually repressing genetic switch \cite{Allen05,Allen06_1}. Details of the FFS method and its implementation are given in the {\em{Supporting Information}}. 

Our results are listed in Table \ref{tab:rates}. In the absence of DNA looping (top two rows of Table \ref{tab:rates}), the lysogen is stable for $\sim 10^9$ generations only when nonspecific DNA binding is absent. When nonspecific binding is included in the model, the lysogen becomes much less stable: a relatively modest nonspecific binding strength $\Delta G_\nsb=-2.1\kcm$ produces a lysogen that is only stable for $\sim 1700$ generations: approximately a million times less stable than the experimental lower bound. In contrast, when DNA looping is included in the model (bottom three rows of Table \ref{tab:rates}), extremely stable lysogens can be achieved  for a wide range of DNA looping  and nonspecific DNA binding strengths. These lysogenic states are in fact even {\em{more}} stable than those observed experimentally \cite{Little99}. One possible explanation might be the effects of passing DNA replication forks (not included in the model), which might be expected to destabilize looped DNA configurations and/or remove bound CI from the operator sites.  The anti-immune state for our model is much less stable than the lysogenic state, in agreement with experimental observations \cite{Calef,Svenningsen05}.

\begin{table}[h]
\caption{Spontaneous switching times (inverse of calculated switching rates) from the lysogenic to anti-immune states and from the anti-immune to lysogenic states, for wild-type phage $\lambda$, computed using FFS.  \label{tab:rates}}
\begin{tabular}{c|c|c|c}
  $\Delta G_\lp$    & $\Delta G_\nsb$ & Switching time   & Switching time  \\
 $\kcm$ &  $\kcm$ & lysogen $\to$ anti-immune & anti-immune  $\to$ lysogen  \\
&&(generations)&(generations)\\
\hline
 no loop  & no NSB   &  $(3.6 \pm 0.1)\times 10^{9}$ &  $2300 \pm 100$   \\
 no loop  & -2.8   & $1700 \pm 50$  & $(7 \pm 1) \times 10^{9}$ \\
\hline
 -5.2  & -4.1   & $< 10^{23}$ & $43 \pm 1$ \\
 -3.7  & -2.8   &  $(3 \pm 2)\times 10^{25}$ &  $26 \pm 1$  \\
 -1.0  & no NSB  &  $(6 \pm 2) \times 10^{14}$  & $130 \pm 10$ \tablenote{A cell generation time of 34 min is assumed. FFS calculations used  8-85 interfaces, 1000-10000 configurations at the first interface and 500-10000 trials per interface, and were  averaged over 10-100 runs.}  
\end{tabular}
\end{table}

\section{DNA looping allows the switch to function despite CI depletion}

It is believed that a large fraction of the transcription factors in a
bacterial cell are unavailable for binding to their specific binding
sites because they are nonspecifically bound to genomic DNA \cite{Hippel74}. 
Recent expression measurements for the $P_{RM}$
promoter have suggested that this is the case for CI
\cite{Dodd01,Dodd04,Bakk04_1}. This nonspecific binding  poses a severe
challenge to the phage $\lambda$ switch. Although both CI and Cro are depleted by nonspecific binding, the $P_{RM}$ promoter
is intrinsically weak, and requires activation by CI at $O_R2$ to
compete effectively with $P_R$. One would therefore expect nonspecific DNA
binding by CI to drastically destabilize the lysogen, and to
compromise the  bistability of the switch. Table \ref{tab:depletion} of the {\em{Supporting Information}} shows how the concentration of free CI depends on the nonspecific DNA binding strength. To characterise the effects on switch function, we investigated the range
of nonspecific binding strengths  over which our
model gave bistability, for different values of the DNA looping
parameter $\Delta G_{\lp}$. Our results are shown in Figure
\ref{fig:phdiag}. In the absence of DNA looping
(top row of Figure \ref{fig:phdiag}), bistability is indeed strongly
compromised as the parameter $\Delta G_{\nsb}$ is increased in
magnitude (left to right). For $|\Delta G_{\nsb}|>\sim 3 \kcm$, the model
is no longer bistable; the lysogenic state cannot be sustained and
only the anti-immune state is stable. However, when the DNA looping
interaction is included in the model, bistability is maintained over a
much wider range of  $\Delta G_{\nsb}$ values. In figure
\ref{fig:phdiag}, the width of the green (bistable) region increases
dramatically as the parameter $|\Delta G_{\lp}|$ increases.  Our model
therefore suggests that one role of the DNA looping interaction may be
to ensure that the switch continues to function even when the level of
intracellular free CI is depleted by nonspecific DNA binding
\cite{Dodd01,Dodd04} or by cell-to-cell fluctuations \cite{Baek03}. We
note that this stability to CI fluctuations is not expected to prevent
lysis from occurring on UV irradiation of the wild-type phage: even for
a strong looping interaction, rapid degradation of CI by RecA will
eventually lead to so little CI being present that the loop cannot be
maintained, upon which lysis will occur. To check this, we simulated a version of the model in which we fixed the total CI concentration, for a typical parameter set ($\Delta G_\lp = -3.7 \kcm$, $\Delta G_\nsb = -2.8 \kcm$). When we artificially lower the CI concentration to $\sim 10 \%$ of the lysogenic steady-state level (30 CI molecules per cell), the system flips to the anti-immune state. This result is in good agreement with the observation of Bailone {\em{et al}} \cite{Bailone1979} that prophage induction occurs at a CI concentration about 10$\%$ of that in the lysogen.

\section{DNA looping causes robustness to operator mutations}

In an important series of experiments, Little {\em{et al}} showed that
the basic functions of the phage $\lambda$ regulatory network are
robust to changes in network architecture. Mutants $O_R(121)$ and
$O_R(323)$, in which the order of the $O_R1$, $O_R2$ and $O_R3$
binding sites was altered compared to the wild-type $O_R(321)$, formed
stable lysogens (although less stable than the wild-type) which could
be induced to enter the lytic pathway on UV irradiation \cite{Little99}. To our knowledge, this robustness has not been reproduced in
computer models \cite{Aurell02}.

\begin{figure}[!h]
\includegraphics[width=.55\textwidth,clip=true]{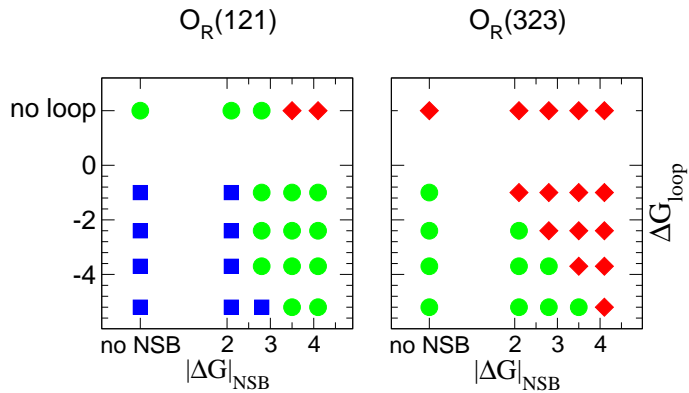}
\caption{\label{fig:phdiag_little} Range of bistability of the model switch as a function of the nonspecific DNA binding parameters $\Delta G_{\nsb}$ and $\Delta G_{\lp}$, in the presence and absence of the DNA looping interaction, for the Little mutants. Symbols have the same meaning as in Figure \ref{fig:phdiag}.}
\end{figure}

We tested whether our model was able to produce stable lysogens for the $O_R(121)$ and $O_R(323)$ mutants, which were created in our simulations by changing the operator binding site affinities. We neglect possible changes in the properties of  $P_{RM}$ \cite{hammer2006}, whose DNA sequence is also affected by the Little {\em{et al}} substitutions. The range of parameters for which our model mutants are bistable is shown in Figure \ref{fig:phdiag_little}.
We find that the DNA looping interaction plays a key role in ensuring robustness.  In the absence of DNA looping (top row of Figure \ref{fig:phdiag_little}), our model produces a stable lysogen for $O_R(121)$ only if the nonspecific binding strength $\Delta G_\nsb$ is below a critical value, and cannot sustain a stable lysogen for $O_R(323)$ at all.  However, when DNA looping  is included in the model (lower rows of Figure \ref{fig:phdiag_little}), the  $O_R(323)$ mutant can achieve a stable lysogenic state, and the $O_R(121)$ lysogen becomes able to tolerate stronger nonspecific DNA binding interactions. In fact, if the looping strength is too strong, our model predicts that the  $O_R(121)$ mutant may lose the ability to sustain a stable anti-immune state.  

The steady-state concentrations of CI in the lysogen predicted for these mutants are in reasonable agreement with the observations of Little {\em{et al}}. For a typical parameter set ($\Delta G_\lp = -3.7 \kcm$, $\Delta G_\nsb = -2.8 \kcm$), we obtained steady-state lysogenic CI concentrations $38\%$ and $81\%$ of the wild-type values for $O_R(121)$ and $O_R(323)$ respectively, in comparison to $25$--$30\%$ and $60$--$75\%$ measured by Little {\em{et al}}. Calculation of the spontaneous switching rates for the $O_R(121)$ and
$O_R(323)$ mutants, using FFS, also shows results qualitatively in agreement
with those of Little {\em{et al}} \cite{Little99} (see table of
results in the {\em{Supporting Information}}). For all combinations of
$\Delta G_\lp$ and $\Delta G_\nsb$, the $O_R(323)$ lysogen is less
stable than $O_R(121)$ which is in turn less stable than the
wild-type, although quantitatively the magnitude of this destabilization is greater than the
factors of 10 observed by Little {\em{et al}}. Our model allows us to make the tentative prediction that (in a $recA$-- host) the stability of the anti-immune states will be $O_R(323) > $
wild-type $> O_R(121)$. These stabilities have not yet been measured \cite{littlepc}.

\section{Alternatives to looping}
Our results indicate that the DNA looping interaction allows the phage
$\lambda$ switch to maintain an extremely stable lysogenic state,
which is robust to operator mutations, while retaining its essential
bistability. Could this result have been achieved by evolution in any other
way? Recent work by Babi{\'{c}} and Little \cite{Babic07} has shown
that a triple mutant lacking CI cooperativity but with increased
$P_{RM}$ activity and enhanced binding of CI to $O_R2$, can maintain a
stable lysogen, with increased CI levels compared to the wild-type,
and can switch into the lytic state. It is very unlikely that this mutant can form a DNA loop \cite{littlepc}. This suggests that increased lysogenic CI levels could substitute for cooperative interactions (including DNA looping). To test whether our model could reproduce these experiments, we modified our simulation
parameters to represent the two triple mutants $\lambda$ $cI$ Y210N
{\em{prm252}} $O_R2up$ and $\lambda$ $cI$ Y210N {\em{prmup-1}}
$O_R2up$. In both cases, we removed all cooperative interactions
between CI dimers, including looping, and increased the affinity of CI
for $O_R2$ by a factor of 5. For the {\em{prm252}} mutation, the basal
and stimulated $P_{RM}$ transcription rates were increased by factors
of 6 and 2.5 respectively, and for the {\em{prmup-1}} mutation these
factors were 10 and 5 \cite{Babic07}. With nonspecific binding
strength $\Delta G_{\nsb}=-2.1 \kcm$, both these ``mutants'' formed
stable lysogens, with CI levels $\sim 3$ and $\sim 6$ times that of
the ``wild-type''. The mutant representing $\lambda$ $cI$ Y210N
{\em{prmup-1}} $O_R2up$ also formed a stable lysogen for $\Delta
G_{\nsb}=-2.8 \kcm$, but $\lambda$ $cI$ Y210N {\em{prm252}} $O_R2up$
did not. These results are in reasonable agreement with those reported by Babi{\'{c}} and Little \cite{Babic07}. Furthermore, our results support the view that  increased lysogenic CI levels could provide an alternative mechanism for maintaining a stable lysogenic state, although this mechanism has apparently not been selected by evolution.

\section{Discussion}
The phage $\lambda$ switch represents an important test for computational modeling of gene regulatory networks. This network has been a paradigm in molecular biology for many years: its architecture and biochemical parameters have been extensively studied and its behavior has been thoroughly and quantitatively characterized. Nevertheless, computational models have failed to produce results in agreement with experimental observations. If modeling is to prove a useful tool in the analysis of larger and more complex gene regulatory networks, it is essential that it should  produce convincing results for phage $\lambda$. 

The computer simulation model presented here reproduces for the first
time both the bistability of the phage $\lambda$ switch and its
extremely low spontaneous flipping rate. Our model includes only known
biochemical interactions and uses as far as possible biochemically
measured parameters. The forward flux sampling (FFS) rare event
sampling method allows us to quantify the stability of the lysogenic
and anti-immune states, even though spontaneous switching rates would
never be observed in a typical brute-force simulation run. An
important difference between our model and previous theoretical
studies is that we simulate the full transcription factor-DNA binding
dynamics. These ``operator state fluctuations'' were eliminated in previous models, using the physical-chemical quasi-equilibrium assumption of Shea and Ackers \cite{Aurell02,Santillan04,Shea85}. Our previous work \cite{Morelli2008_2}, as well as that of Walczak and Wolynes \cite{Walczak05}, has shown that operator state fluctuations can drastically affect the switching pathways and the switching rate for bistable genetic networks. Moreover, Vilar and Leibler \cite{Vilar03} have shown that operator state fluctuations are coupled to DNA looping dynamics. We therefore model explicitly both the protein-DNA binding dynamics and the DNA looping dynamics. Our model also includes nonspecific DNA binding, as does ref. \cite{Aurell02}. Although DNA looping and nonspecific binding are modeled in a simplified manner, our model involves many reactions and is computationally expensive to simulate. This problem is alleviated by the FFS method, in combination with the coarse-graining of dimerization and non-specific binding reactions. We note that a semi-analytical approach, based on Large Deviation Theory, which has been applied successfully to a simplified model of the phage $\lambda$ switch  \cite{roma}, may provide a promising alternative to direct simulations as performed here.

Our results suggest a key role for the DNA looping interaction in
ensuring that the network retains its bistability even when exposed to
perturbations. Our model requires DNA looping to achieve bistability
in the presence of nonspecific DNA binding and/or operator state
mutations. The $P_{RM}$ promoter is intrinsically weak and requires CI
to be bound at $O_R2$ in order to compete effectively with $P_R$. The
DNA loop enhances CI occupancy at $O_R2$, providing a way to achieve
sustained activation of the $P_{RM}$ promoter, even when the free
intracellular CI concentration is very small. We also find, in
agreement with Dodd {\em{et al}}, that the presence of the loop
increases autorepression of $P_{RM}$ by enhancing CI binding at
$O_R3$ \cite{Dodd01}. Looping thus increases the strength of both
auto-activation via $O_R2$ and auto-repression via $O_R3$; this
reduces the fluctuations in CI, which enhances the stability of
the switch \cite{Warren06}. Our model does not include the loop-mediated increase in
maximal $P_{RM}$ activity recently discovered by Anderson and Yang
\cite{anderson2008}; we expect that including this in the model would
only strengthen our conclusions.

Our simulations and the experiments of Babi{\'{c}} and Little \cite{Babic07}
show that a stable lysogen can be obtained in the absence of
DNA looping and other cooperative interactions, by raising the level of CI. This observation is perhaps not so surprising, since the stability of genetic switches is predicted to depend exponentially on the expression
levels of gene regulatory proteins \cite{Warren06,Warren04}. What is then the role of DNA looping? DNA looping not only reduces fluctuations in CI levels, as discussed above, but also reduces operator state fluctuations, which have been predicted to limit the stability of genetic switches \cite{Allen05,Morelli2008_2,Walczak05}. This suggests that the looping interaction allows a stable lysogen to be achieved with lower CI levels. One advantage of this may be a reduction in the energetic cost to the host cell of producing CI \cite{TanaseNicola08,Dekel05}. 
 Alternatively, the  dynamical pathways for the transition to lysis may be more favorable for a switch with DNA looping.

Several predictions emerge from our simulations. Firstly, we predict that a mutant phage which cannot
form the DNA loop will form a lysogen with much reduced stability
compared to the wild-type, and may not be able to sustain a lysogen at
all. We further predict that depleting free CI (for example by
introducing a large number of specific binding sites) will strongly
destabilize the lysogen in a non-looping mutant but will have a much
less severe effect on the wild-type lysogen. Finally, our simulations allow us to tentatively suggest an order for the stability of the anti-immune states in a non-lysing version (for example $O_R(121) N^-O^-$) of the Little mutants in a  $recA$-- host.

This work presents an encouraging picture of the potential for
computational modeling to unravel the contribution of different
biochemical mechanisms to observed biological behavior. Using a
simplified representation of the core part of the phage $\lambda$
switch,  we are able
to obtain behavior in qualitative and partial quantitative agreement
with experimental results. Future work should include more
sophisticated models for DNA looping and nonspecific binding, explicit
modeling of cell growth, DNA replication and cell division, as well
as detailed characterization of the switching pathways. Such work
should make stochastic modeling, in combination with experiments, an
important tool in unraveling the mechanisms behind the complex
behavior of biochemical networks.

\begin{acknowledgments}
The authors are very grateful to Ian Dodd, John Little and Noreen Murray for discussions and advice, and also to Andrew Coulson,  David Dryden, Andrew Free, Davide Marenduzzo and Sorin T{\u{a}}nase-Nicola for assistance during the course of this work. This work is part of the research program of the
"Stichting voor Fundamenteel Onderzoek der Materie (FOM)", which is financially supported by the "Nederlandse organisatie voor
Wetenschappelijk Onderzoek (NWO)''.  M.J.M. acknowledges financial support from the European Commission, under the HPC-Europa programme, contract number RII3-CT-2003-506079. R.J.A. was funded by the Royal Society of Edinburgh.
\end{acknowledgments}

\section{SUPPLEMENTARY MATERIAL}

In section \ref{sec:sim_dets}, we discuss in more detail the technical aspects of our stochastic simulations. In section \ref{sec:sim_res}, we present results including steady-state protein concentrations, depletion of free CI by nonspecific binding, promoter activity as a function of CI concentration, switching rate calculations for the Little mutants and effects of increasing the RNAp concentration. In section \ref{sec:params}, we test the sensitivity of our conclusions to variations in the parameters of the model. Finally, in Appendix \ref{sec:lists}, we give a complete list of the chemical components and reactions in our model. We note that many of these reactions are permutations of each other, so that the number of rate constant parameters is much smaller than the number of reactions.

\section{Simulation Details}\label{sec:sim_dets}

In section \ref{sec:loop} we discuss how DNA looping is represented in our model. In sections \ref{sec:cgdim} and  \ref{sec:cgdna} we give details of how we coarse-grain the dimerization reactions and the nonspecific DNA binding reactions in the model. In section \ref{sec:ffs} we discuss Forward Flux Sampling and its application to this model.

\subsection{Modeling DNA looping}\label{sec:loop}

\subsubsection*{Representation of looping in our simulation model}

In our model, the  $O_L$ operator has three binding sites,  $O_L1$, $O_L2$ and $O_L3$, which can bind CI or Cro dimers, as well as a promoter, $P_L$, which overlaps the binding site $O_L1$, and which can bind RNA polymerase. It is important to include RNAp binding to $P_L$, because this  excludes CI binding to $O_L1$ (even though no gene is expressed from $P_L$ in the model). Binding constants for Cro dimers to the three $O_L$ binding sites are assumed to be the same as for $O_R$. For CI, we use the binding constants measured by Dodd {\em{et al}} \cite{Dodd04}. 

In our model, there are 3 copies of each of the $O_R$ and $O_L$ operators in the cell (since we assume a doubling time of 34 minutes). We assume that any of the copies of $O_R$ can form a loop with any copy of $O_L$, since the 2400bp separation between $O_R$ and $O_L$ on the $\lambda$ genome is long compared with the persistence length of DNA ($\sim 50\nm = 150$bp). The system can form a loop if both the $O_R$ and $O_L$ operators are bound by at least two adjacent pairs of CI dimers ({\em{i.e.}} in the 1-2 or 2-3 positions). Fully bound operators (with dimers in all 3 positions) can also form loops. We classify looped states into four categories, according to which binding sites are involved in octamer formation. Loops where the octamer is formed by CI dimers at   $O_R1$, $O_R2$, $O_L1$ and $O_L2$ are denoted $OLR1$, loops whose octamer is formed by dimers at $O_R1$, $O_R2$, $O_L2$ and $O_L3$ are denoted $OLR2$, loops whose octamer is formed by dimers at $O_R2$, $O_R3$, $O_L1$ and $O_L2$ are denoted $OLR3$ finally loops with octamers formed by CI dimers at $O_R2$, $O_R3$, $O_L2$ and $O_L3$ are denoted $OLR4$. For each of these categories, the additional two binding sites, not involved in octamer formation, may also be occupied by CI or Cro dimers (or in some cases RNAp). We therefore designate any particular looped state by $OLRX(YZ)$, where X is 1, 2, 3 or 4, as above, and Y and Z denote the species that are bound to the two non-octamerized sites on $O_L$ and $O_R$ respectively. For example, a looped state where CI dimers bound at $O_R1$, $O_R2$, $O_L1$ and $O_L2$ participate in an octamer, and additional CI dimers are bound at $O_R3$ and $O_L3$ would be denoted as $OLR1(RR)$ in the reaction scheme in Appendix \ref{sec:lists} of the {\em{Supporting Information}} (here, ``R'' denotes ``CI repressor''). 

Loop formation is represented in our reaction scheme (see Appendix \ref{sec:lists}) by reactions in which unlooped states with CI-bound  $O_R$ and $O_L$ operators associate to and dissociate from the appropriate $OLR$ states. Association to a looped state always occurs with rate $k_\lp$, and dissociation always occurs with rate $k_\ulp$, regardless of which combination of CI dimers and/or RNAp is bound. If a looped state has vacant binding sites, binding to these is assumed to be  diffusion-limited. To take account of the cooperative binding interaction due to tetramer formation between the two non-octamerized CI dimers in the fully occupied looped state \cite{Dodd04}, we reduce the dissociation rate of a single dimer from the fully occupied looped states $OLR1(RR)$ and $OLR4(RR)$ by a factor  $\exp{[-\Delta G_{\tet}/RT]}$ where $\Delta G_{\tet}=-3 \kcm$ \cite{Dodd04}.  This scheme for including the cooperativity due to tetramerization ignores the possibility that loops may form in either orientation \cite{anderson2008}, but we believe that this simplification is likely to have only a minor effect on our results.

\subsubsection*{Estimation of DNA looping parameters from polymer dynamics}

The strength of the DNA loop between $O_R$ and $O_L$ {\em{in vivo}} is
unknown. We therefore vary the  ratio $k_\lp/k_\ulp \equiv \exp{[-\Delta G_{\lp}/RT]}$ in our simulations, maintaining a fixed value of  $62.1\scc^{-1}$ for $k_\lp$ and varying $k_\ulp$ [as for protein-DNA binding, we assume that the binding affinity affects only the ``off''-rate]. In some cases, where this would result in extremely slow loop dissociation (slower than the cell doubling time), we increase $k_\lp$ to prevent the looping dynamics becoming very slow.  In all the cases we have examined, we find that only the ratio $k_\lp/k_\ulp$ is important, indicating that in our model, DNA looping/unlooping is not the rate limiting step for switch flipping. 

Although we use $\Delta G_{\lp}$, and hence $k_\ulp$, as an adjustable parameter in our simulations, we can gain some insight into likely values for $k_\lp$ from a consideration of polymer dynamics. For simplicity, we assume that the association time  for two CI-bound operators, separated by 2400bp, to form a loop, is of the same order of magnitude as the mean relaxation time of a 2400bp DNA molecule, whose ends have been brought into contact. This is given by Meiners as \cite{Meiners2000}:
\begin{equation}
t_a=\frac{4\eta L_0^2 l_p}{\pi k_bT\ln(L_0/d)}
\end{equation}
 In this expression, $\eta$ is the viscosity coefficient 
for the cytoplasm, $L_0=2400bp=800nm$ is the contour length of the DNA, $l_p$ is the persistence
length and $d=2.5nm$ is the thickness of DNA. Assuming the viscosity of the cytoplasm to be ten times that of water \cite{kasza07}, $\eta \approx 10 \times 10^{-3}$ Pa.s, we obtain $t_a=0.016\scc$. This leads to  $k_\lp = t_a^{-1} = 62.1\scc$, which is the typical value used in our simulations. We then vary  $k_\ulp$ according to the chosen value of $\Delta G_{\lp}$.

In our simulations, we vary $\Delta G_\lp$ over a range larger than the value of $-0.5 \kcm$ obtained by Dodd {\em{et al}} by fitting promoter activity data for $P_{RM}$ and $P_R$. We assess the extent to which our model fits this promoter activity data in Section \ref{sec:pa}. As discussed by Dodd {\em{et al}} \cite{Dodd04}, the true value of  $\Delta G_\lp$ {\em{in vivo}} is unknown. {\em{In vitro}} measurements of the free energy of association of two CI tetramers in the absence of DNA give a value of $-9.1\kcm$ \cite{senear93}, but this may be different in the presence of DNA, and in any case must be offset by a contribution due to the entropy of DNA looping.

\subsection{Coarse-graining dimerization}\label{sec:cgdim}
The monomer-dimer association and dissociation reactions: 
\begin{eqnarray}\label{ass_diss}
\ci+\ci &\rightleftharpoons \ci_2\\ 
\nonumber \cro+\cro &\rightleftharpoons \cro_2
\end{eqnarray}
are responsible for the  vast majority ($\sim 99 \%$) of the total computational effort when our model switch is simulated in the absence of looping and nonspecific DNA binding. This is because these reactions have very high propensities, due to the large number of CI/Cro molecules in the lysogenic/anti-immune stable states. We can greatly increase the efficiency of our simulations, without jeopardizing their accuracy, by ``coarse-graining'' these reactions. We assume that the monomer-dimer association and dissociation reactions are fast enough that they reach steady state on the timescale of the other, slower reactions in the system (although there is some evidence that Cro dimerization kinetics may be slow \cite{jia05}). We can then remove these reactions from our reaction set \cite{Bundschuh03,Cao05,morelli2008}, as in our previous work on a simpler genetic switch model \cite{morelli2008,morelli2008_2}. To do this, we define new simulation variables:
\begin{eqnarray}\label{ed1:defs}
n_{\check{\ci}} & \equiv & n_\ci+2n_{\ci_2}\\
\nonumber n_{\check{\cro}} & \equiv & n_\cro+2n_{\cro_2}
\end{eqnarray}
where $n_\ci$ represents the number of CI monomers, $n_{\ci_2}$ the number of CI dimers and  $n_{\check{\ci}}$ the sum of these numbers (and likewise for Cro). We rewrite our reaction scheme in terms of the new ``chemical species''  $\check{\ci}$ and $\check{\cro}$, whose numbers remain unchanged by the association/dissociation reactions (\ref{ass_diss}). Translation reactions now produce $\check{\ci}$ or $\check{\cro}$, rather than CI or Cro, but with propensity computed as before. Protein removal from the system by dilution now removes $\check{\ci}$ or $\check{\cro}$, also with reaction propensity unchanged. The active degradation of Cro monomers now becomes degradation of $\check{\cro}$. Here, the new reaction propensity is given by the old rate constant, multiplied by the average number $\langle n_\cro \rangle_{n_{\check{\cro}}}$ of Cro monomers that would be obtained by a simulation of the fast association/dissociation reactions, at fixed  $n_{\check{\cro}}$ [in what follows, we will use angular brackets with a subscript to denote an average value of the quantity in the brackets, taken over a simulation in which the subscripted quantity is held constant]. All dimer-DNA dissociation reactions now produce $\check{\ci}$ or $\check{\cro}$ instead of $\ci_2$ or $\cro_2$, with propensity computed as in the full reaction scheme. All dimer-DNA association reactions are now represented by the association of two units of  $\check{\ci}$ or $\check{\cro}$  to $O_R$ or $O_L$, with propensities equal to the original reaction rate, multiplied by the average number $\langle n_{\ci_2} \rangle_{n_{\check{\ci}}}$ or $\langle n_{\cro_2} \rangle_{n_{\check{\cro}}}$ of dimers given by the association/dissociation reaction set (\ref{ass_diss}) for fixed  $n_{\check{\ci}}$ or  $n_{\check{\cro}}$. To implement this scheme, we need to know $\langle n_{\ci} \rangle_{n_{\check{\ci}}}$,  $\langle n_{\cro} \rangle_{n_{\check{\cro}}}$, $\langle n_{\ci_2} \rangle_{n_{\check{\ci}}}$ and $\langle n_{\cro_2} \rangle_{n_{\check{\cro}}}$, for all possible values of $\check{\ci}$ and $\check{\cro}$. These values are obtained prior to the main simulation run, by numerical solution of the chemical master equations corresponding to Eq.(\ref{ass_diss}), for fixed $\check{\ci}$ and $\check{\cro}$ \cite{morelli2008}. The results are stored in a table for all values of $\check{\ci}$ and $\check{\cro}$. This table is then used to compute the necessary reaction propensities in our simulations of the coarse-grained phage $\lambda$ reaction set.

We have demonstrated this procedure \cite{Bundschuh03,Cao05} in previous work on a model bistable switch formed of two mutually repressing genes \cite{morelli2008}. In that work, we showed that coarse-graining the monomer-dimer association/dissociation reactions had little effect on the switch flipping rate - in contrast to the dimer-DNA association/dissociation reactions, which could not safely be coarse-grained. We have also verified that for our phage $\lambda$ model (in the absence of DNA looping or nonspecific binding, for simplicity), the lysogenic to anti-immune switching rate computed with explicit monomer-dimer association/dissociation is indistinguishable from the same rate as computed with the coarse-grained reaction set. 

\subsection{Coarse-graining nonspecific DNA binding}\label{sec:cgdna} 

Nonspecific binding of CI and Cro dimers to {\em{E. coli}} genomic DNA is included in our model via a set of association-dissociation reactions to genomic DNA sites (``D''), of which we assume there are $10^7$ (about 70\% of the total base pairs in 
the three copies of the {\em{E. coli}} genome that we suppose to be present). These reactions are:
\begin{eqnarray}\label{ass_diss1}
\ci_2+\dd &\rightleftharpoons& \dd\ci_2\\ 
\nonumber \cro_2+\dd &\rightleftharpoons& \dd\cro_2
\end{eqnarray}
These reactions do not necessarily represent the actual cellular process which results in depletion of free intracellular CI and Cro. Such processes are likely to be very complex: these reactions simply provide a convenient and simple way to deplete the available CI in our simulations. The nonspecifically bound species DCI$_{\rm{2}}$ and   DCro$_{\rm{2}}$ are assumed to be subject to dilution due to cell growth, which we model by removing them from the system with rate constant $\ln(2)/T$, where $T=34$ min is the cell cycle time. The free energy $\Delta G_\nsb$ of binding of a CI or Cro dimers to a D site is varied systematically in our simulations. We assume this parameter takes the same value for CI and Cro. The association rate $k_a$ is assumed to
be diffusion-limited, and the dissociation rate $k_{d}$ is calculated as described in the main text. However, due to the large number of D sites, the propensities for these reactions are very much larger than those for any other reactions in the system. Direct Gillespie simulation of these reactions is not computationally feasible. Instead, we use a coarse-graining scheme which we construct according to the principles described above for the dimerization reactions, and in reference \cite{morelli2008}. In this scheme, we coarse-grain both the dimerization and nonspecific binding reactions. This implies that we assume that both association/dissociation to D sites, and association/dissociation of dimers, occur on much faster timescales than the other reactions in the system.

To carry out this coarse-graining, we define new simulation variables:

\begin{eqnarray}\label{ed2:defs}
n_{{\ci'}} & \equiv & n_\ci+2n_{\ci_2}+2n_{\dd\ci_2}\\
\nonumber n_{{\cro'}} & \equiv & n_\cro+2n_{\cro_2}+2n_{\dd\cro_2}
\end{eqnarray}

which represent the total number of CI and Cro molecules in the system, excluding only those which are specifically bound to the operators $O_R$ and $O_L$. Our reaction scheme in fact remains exactly the same as that described in section \ref{sec:cgdim} above, except that $\check{\ci}$ and $\check{\cro}$ are replaced by $\ci'$ and $\cro'$, and the averages $\langle n_{\ci} \rangle_{n_{\check{\ci}}}$,  $\langle n_{\cro} \rangle_{n_{\check{\cro}}}$, $\langle n_{\ci_2} \rangle_{n_{\check{\ci}}}$ and $\langle n_{\cro_2} \rangle_{n_{\check{\cro}}}$ are replaced by $\langle n_{\ci} \rangle_{n_{{\ci'}}}$,  $\langle n_{\cro} \rangle_{n_{{\cro'}}}$, $\langle n_{\ci_2} \rangle_{n_{{\ci'}}}$ and $\langle n_{\cro_2} \rangle_{n_{{\cro'}}}$. These latter averages must be computed over both the dimerization reaction set (\ref{ass_diss}) and the nonspecific binding reaction set (\ref{ass_diss1}), for fixed values of $n_{\ci'}$ and $n_{\cro'}$. We assume that the reaction sets for CI and Cro do not couple ({\em{i.e.}} that the number of D sites is very large), so that we can compute the two sets of averages independently. We carry out a preliminary set of simulations, in which only reactions (\ref{ass_diss}) and (\ref{ass_diss1}) are simulated, for either CI or Cro, for fixed $n_{\ci'}$ or $n_{\cro'}$. Using these simulations, we evaluate the necessary averages, which we tabulate for all values of $n_{\ci'}$ and $n_{\cro'}$. We then use these tabulated values to compute the required propensities in our coarse-grained simulations of the phage $\lambda$ switch. This coarse-graining leads to a speedup of our simulations by a factor of at least 100. Without this coarse-graining, the computations described here would have been beyond the limits of our computational resources.

\subsection{Forward flux sampling for the phage $\lambda$ switch}\label{sec:ffs}
\begin{figure}[!h]
\includegraphics[width=.6\textwidth,clip=true]{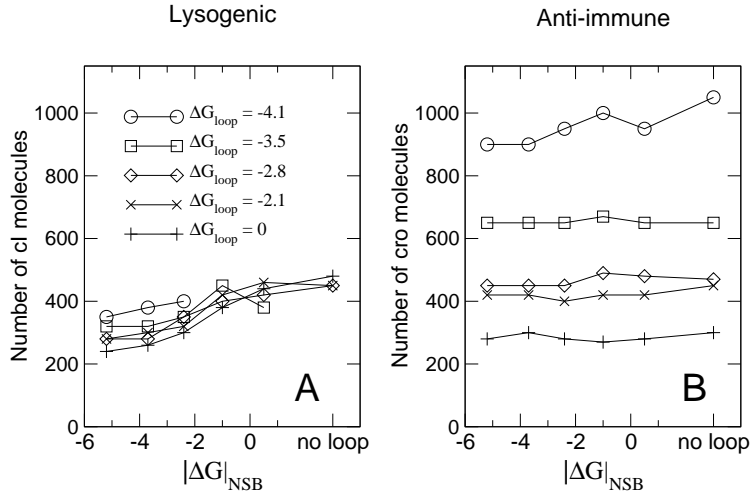}
\caption{Steady-state total number of molecules of CI in the lysogenic state and of Cro in the anti-immune state, as a function of the DNA looping strength $\Delta G_{\lp}$, for different values of the nonspecific binding strength $\Delta G_{\nsb}$. The lysogenic CI concentration is in agreement with measured values and is rather insensitive to nonspecific binding, although DNA looping causes a decrease in the lysogenic CI concentration. The concentration of Cro in the anti-immune state is insensitive to DNA looping strength (as expected) but increases with the strength of the nonspecific binding interaction.\label{fig:S1}}
\end{figure}

Forward Flux Sampling (FFS) allows the calculation of rate constants and transition paths for rare events in equilibrium and nonequilibrium stochastic simulations. FFS is described in detail in refs \cite{Allen05} and \cite{Allen06_1}. It will be sketched briefly here in the context of the transition from the lysogenic to anti-immune states of our model. We define  a collective variable of the system, or order parameter, that distinguishes the initial and final states. We do not assume that this order parameter is the true ``reaction coordinate'' for the transition.  For the lysogenic to anti-immune transition, our order parameter $N$ is taken to be the difference between the total numbers of CI and Cro molecules in the system:
\begin{equation}
N={\mathrm{total~number~of~Cro}}-{\mathrm{total~number~of~CI}}
\end{equation}
In the anti-immune state, Cro dominates and $N$ is positive. In the lysogen, CI dominates and $N$ is negative. We define values $N_\lys$ and $N_\ai$ such that if $N<N_\lys$, the system is assumed to be in the lysogenic stable state, and if $N>N_\ai$, the system is assumed to be in the anti-immune stable state. The chosen values of $N_\lys$ and $N_\ai$ vary for different parameter values, since the numbers of CI and Cro in the two stable states are affected by the choice of parameters (as shown in Figure \ref{fig:S1}). The value of $N_\lys$ should be such that, when the system is in the steady state, fluctuations quite frequently take it into the region where $N>N_\lys$.

We note that although we have chosen $N$ as our order parameter, we could have made an alternative choice - for example, the total number of CI molecules, or some weighted combination of the numbers of CI and Cro. Providing these other choices give good separation of the initial and final states, the same rate constants and transition paths should be obtained. However, the computational efficiency and sampling effectiveness may be affected by the choice of order parameter.

The aim of the computation is to compute the rate $k$ at which the system makes transitions from the region where $N<N_\lys$ to the region where $N>N_\ai$. This is the frequency with which simulation trajectories leave the lysogenic state and enter the anti-immune state. This frequency is very low, but it can be re-expressed as \cite{TIS}:
\begin{equation}\label{tis1}
k=\Phi_\lys P_{\lys \to \ai}
\end{equation}
Here, $\Phi_\lys$ is the flux of trajectories out of the lysogenic state, or the frequency with which the system crosses the interface $N=N_\lys$ coming from the lysogenic state. This flux must be multiplied by the probability $P_{\lys \to \ai}$ that one of these trajectories which leaves the lysogenic state will subsequently reach the anti-immune state rather than returning to the lysogenic state. The flux $\Phi_\lys$ can be calculated very accurately because the system makes many crossings of $N=N_\lys$. The probability $P_{\lys \to \ai}$ is however very small and difficult to calculate. This problem can be overcome by placing a number of ``interfaces'' between the lysogenic and lytic states, at values of $N$ intermediate between $N_\lys$ and $N_\ai$. We define $N_0=N_\lys$, $N_n=N_\ai$, and place interfaces $N_i$ for $1 \le i \le n-1$ between $N_0$ and $N_n$. Eq. (\ref{tis1}) can then be rewritten as \cite{TIS}: 
\begin{equation}\label{tis2}
k=\Phi_\lys \prod_{i=0}^{n-1} P({N_{i+1}|N_{i}}) 
\end{equation}
where $P({N_{i+1}|N_{i}})$ is the probability that a trajectory which has reached interface $N_i$, coming from the lysogenic state, will subsequently reach the next interface $N_{i+1}$, rather than returning to the lysogenic state. Since the interfaces may be placed arbitrarily close together, the probabilities $P({N_{i+1}|N_{i}})$ can be large, and thus easy to compute accurately. Once these probabilities and the flux $\Phi_\lys$ have been computed, the rate for the lysogenic to anti-immune transition can be obtained using Eq.(\ref{tis2}).

We compute the flux $\Phi_\lys$ using a simulation run initiated in the lysogenic state. Every time the simulation crosses $N=N_\lys$ in the direction of increasing $N$, we increment a counter. At the same time, we store the configuration of the system at the moment that the crossing occurs.  At the end of this simulation run, we divide the value of our counter by the total simulation time to obtain $\Phi_\lys$. We have also obtained a collection of system configurations corresponding to crossings of $N=N_\lys$. 

We now turn to the computation of the probabilities $ P({N_{i+1}|N_{i}})$. To compute $ P({N_{1}|N_{0}})$, we fire many ``trial runs'' from the collection of configurations that we have obtained at $N_0$. In each trial run, we choose a configuration at random from our collection and use it as the starting point for a new simulation run, which is continued until either the system reaches $N_1$, or it returns to the lysogenic state $N<N_0$. If $N_1$ is reached, the final configuration of the trial run is stored in a new collection. We repeat this trial run procedure many times, and finally estimate $ P({N_{1}|N_{0}})$ as the fraction of trial runs that successfully reached $N_1$. We now use the collection of configurations which we have just collected at  $N_1$ to compute $ P({N_{2}|N_{1}})$: once again we fire many trial runs which are continued until either $N_2$ or $N_0$ is reached. We repeat this procedure for each interface, until we have computed all the $ P({N_{i+1}|N_{i}})$ values. At this point, trajectories corresponding to transitions from the lysogenic to anti-immune states could be extracted by tracing back trial runs from $N_\ai$ back to $N_\lys$. An analysis of these trajectories could be used to investigate the transition mechanism. However, in this work we have confined ourselves to computing the rate constant $k$. 

The frequency of transitions from the anti-immune to lysogenic states can be computed in the same way - here, our order parameter is simply taken to be 
\begin{equation}
N'={\mathrm{total~number~of~CI}}-{\mathrm{total~number~of~Cro}}
\end{equation}

To obtain the results presented in this paper, we used 8-85 interfaces, with 1000-10000 configurations collected at $N_0$ and 500-10000 trial runs (precise values varied in each calculation). In addition, we averaged our results over 10-100 FFS calculations, to obtain an estimate for the error bars on our calculated rate constants. The final rate constants obtained are not dependent on the exact values of these parameters \cite{Allen06_1,Allen06_2}. For implementations of FFS with very few ($\sim 10-50$) configurations at each interface, sampling errors may arise if the configurations most likely to progress to the final state are  missed, especially if the order parameter poorly represents the progress of the transition \cite{sear,bolhuis}. These can be detected by the appearance of large variations between the rate constants calculated in different FFS runs. However, we have found this not to be the case when the chosen order parameter is reliable and the number of configurations is more than about 100, as in this work.

\section{Simulation Results}\label{sec:sim_res}

In this section, we present data on steady-state CI and Cro concentrations in the lysogenic and anti-immune states (\ref{sec:ss}), depletion of free CI by nonspecific DNA binding, promoter activity as a function of lysogenic CI concentration (\ref{sec:pa}), switch stability for the Little mutants, calculated using FSS (\ref{sec:little}) and the effects of increasing the RNAp concentration (\ref{sec:Rp}).

\subsection{Steady-state protein concentrations}\label{sec:ss}

Figure \ref{fig:S1} shows the average number of CI and Cro molecules in our simulations of the lysogenic and anti-immune states respectively, for different values of the parameters $\Delta G_{\lp}$ and $\Delta G_{\nsb}$. The model produces steady-state protein concentrations in agreement with measured values.  The number of CI molecules in a wild-type lysogen has been measured to be 220 \cite{Reichardt71}, but is highly variable \cite{Baek03}. The concentration of Cro in the anti-immune state has been measured as $\sim 400\nM$ \cite{Reinitz90}, so that in our assumed volume of 2$\um^3$ we expect $\sim 500$ molecules. Our model predicts that  the steady-state CI level in the lysogen is rather stable to changes in the nonspecific DNA binding strength $\Delta G_{\nsb}$. However, if the looping interaction is absent, the lysogenic state cannot be sustained when  $\Delta G_{\nsb}$ becomes large.

Increasing the strength of the DNA looping interaction decreases the lysogenic CI concentration by a factor of up to 2, in agreement with the observations of Dodd {\em{et al}} \cite{Dodd04}. In contrast, the level of Cro in the anti-immune state  increases strongly with the strength of the nonspecific binding interaction, although it is insensitive to the DNA looping strength (as the DNA loop does not form in the anti-immune state). $P_R$ is repressed by Cro binding at $O_R1$ or $O_R2$. Depletion of free Cro by nonspecific binding reduces this autorepression, increasing expression of $P_R$, and increasing the total cellular Cro level.

\subsection{Depletion of free transcription factors by nonspecific DNA binding}\label{sec:dep}

\begin{table*}[h]
\begin{center}
\begin{tabular}{|c|c|c|c|c|}
\hline
 $\Delta G_\nsb$ ( $\kcm$)   & free monomers & free dimers & NSB dimers & percentage free CI   \\
\hline
-2.1 &  76 & 75 & 37 & 75\%\\
-2.8  & 52 & 48 & 76 & 49\% \\
-3.5  & 23 & 24 & 115 & 24\%  \\
-4.1  & 18 & 9 & 132 & 12\%  \\
\hline
\end{tabular}
\end{center}
\caption{Partioning of the intracellular CI pool into free monomers, free dimers and nonspecifically bound dimers, for several values of $\Delta G_\nsb$, assuming a total of 300 CI molecules in the cell and ignoring specific DNA binding reactions. The percentage of the CI pool which is free [not DNA-bound] is shown in the right-hand column.} \label{tab:depletion}
\end{table*} 
Table \ref{tab:depletion} shows the partioning of the total CI in the cell into free monomers, free dimers and nonspecifically bound dimers, assuming a total of 300 molecules in the cell and ignoring specific DNA binding. This demonstrates that the amount of free CI is dramatically depleted by nonspecific DNA binding, for the range of interaction strengths considered in this work. 

\vspace{0.5cm}

\subsection{Promoter Activity}\label{sec:pa}

In recent experiments by Dodd {\em{et al}} \cite{Dodd01,Dodd04}, the activity of the $P_{RM}$ and $P_R$ promoters was measured as a function of the intracellular CI concentration (supplied from a plasmid), for constructs with and without the DNA looping interaction \cite{Dodd01,Dodd04}. Fitting the resulting data with a thermodynamic binding model using {\em{in vitro}} measured binding constants showed that  CI binding to $O_R$ was much weaker than expected \cite{Dodd01,Dodd04}. Similar effects are well-known for other promoters, including {\em{lac}} \cite{Hippel74}, and have been attributed to nonspecific binding of transcription factors to genomic DNA. Dodd {\em{ et al}} \cite{Dodd01,Dodd04} and other authors \cite{Bakk04_1,Bakk04_2} fitted the measured promoter activity data with a model including nonspecific DNA binding. Although these promoter activity curves may be distorted due to plasmid copy number variation \cite{littledodd}, the reduction in the apparent free CI concentration, apparently due to nonspecific binding, appears to be real and significant.

\begin{figure}[!h]
\includegraphics[width=.6\textwidth,clip=true]{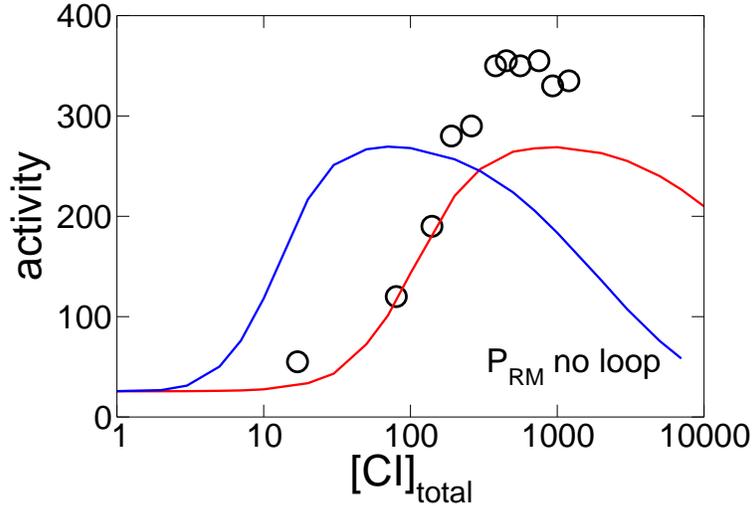}
\caption{Computed activity of the $P_{RM}$ promoter, as a function of total cellular CI concentration. The red lines are computed with $\Delta G_{\nsb}=-4.1 \kcm$; the blue lines without nonspecific DNA binding. The symbols show the data of Dodd {\em{et al}}, 2001 (circles) and Dodd {\em{et al}}, 2004  (squares). (a): $P_{RM}$ activity in absence of DNA looping interaction. \label{fig:S2a}}
\end{figure}

\begin{figure}[!h]
\includegraphics[width=.6\textwidth,clip=true]{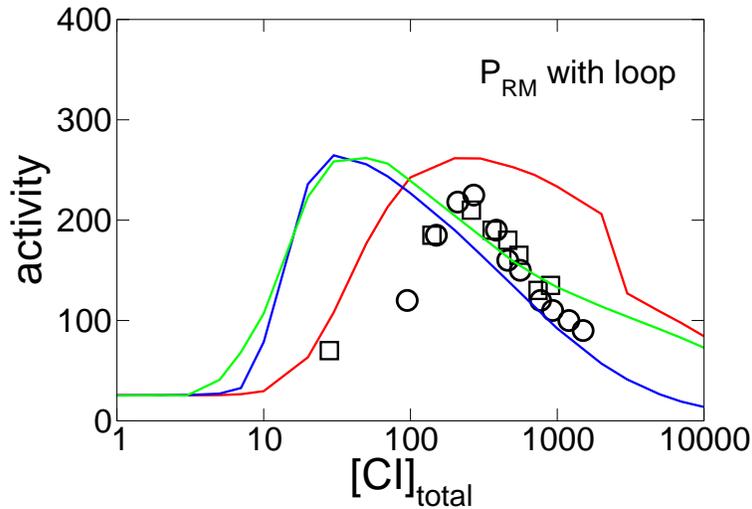}
\caption{Computed activity of the $P_{RM}$ promoter, as a function of total cellular CI concentration. The red lines are computed with $\Delta G_{\nsb}=-4.1 \kcm$; the blue lines without nonspecific DNA binding. The symbols show the data of Dodd {\em{et al}}, 2001 (circles) and Dodd {\em{et al}}, 2004  (squares). $P_{RM}$ activity, with DNA looping. The red and blue lines are computed with  $\Delta G_{\lp}=-1\kcm$. The green line is computed with $\Delta G_{\lp}=-3.7\kcm$ .  The simulation data was multiplied by a scaling factor of 10000. \label{fig:S2b}}
\end{figure}

 In Figures \ref{fig:S2a} and \ref{fig:S2b}, we measure $P_{RM}$ activity as a function of fixed total CI concentration in our model (CI levels are fixed by turning off protein production; no Cro is present). Increasing the nonspecific binding strength (magnitude of $\Delta G_\nsb$)  shifts the promoter activity curves towards higher CI concentrations. The promoter activity data measured by Dodd {\em{et al}}, extracted from Refs \cite{Dodd01} and \cite{Dodd04}, are also shown in Figures \ref{fig:S2a} and \ref{fig:S2b}. To convert Wild-type Lysogenic Units (WLU) to CI concentrations, we assumed a lysogenic CI concentration of 370$\nM$ \cite{Dodd04}. Without nonspecific DNA binding, our model completely fails to agree with the measured $P_{RM}$ activity curves (blue lines).  By fitting their data to a model with strong nonspecific DNA binding [$\Delta G_\nsb = -6.2\kcm$, with $6.76 \times 10^{-3}\M$ nonspecific binding site concentration], Dodd {\em{et al}} obtained a best-fit value of $-0.5 \kcm$ for the DNA looping strength. In our model, the nonspecific binding is weaker compared to Dodd {\em{et al}} ($\Delta G_\nsb \sim 2-4 \kcm$; $1.0 \times 10^7$ sites in a volume of $2\um$ - in the same volume, Dodd {\em{et al}} would have $1.35 \times 10^7$ sites), and the DNA looping interaction is stronger. Dodd {\em{et al}} did not model the dynamics of the system, and therefore were not aiming to reproduce the stability of the lysogenic state. Our model, by contrast, is dynamical. We find that a strong looping interaction is needed to explain the stability and robustness of the lysogenic state; however, with our parameters, we do not fit the  Dodd {\em{et al}} data well, as shown, for example, by the green line in Figure \ref{fig:S2b}. Plasmid copy number variation may account for some of the discrepancy, as may our smaller value of $\Delta G_\nsb$. However, the apparent discrepancy between the large value of $\Delta G_\lp$ needed to obtain agreement with experimental observations in dynamical simulations, and the smaller value obtained from fitting promoter activity data, remains to be resolved.

\subsection{Spontaneous switching times for the Little mutants}\label{sec:little}

\begin{table*}[h]
\begin{center}
\begin{tabular}{|c|c|c|c|c|}
\hline
 Mutant & $\Delta G_\lp$    & $\Delta G_\nsb$ & Switching time   & Switching time  \\
& $\kcm$ &  $\kcm$ & lys. $\to$ anti-imm. & anti-imm.  $\to$ lys.  \\
& &  & (generations) & (generations) \\
\hline
$O_R(121)$ & -5.2  & -4.1   & $(1.1 \pm 0.4) \times 10^{10}$ &  $ 9.5 \pm 0.1$  \\
$O_R(121)$ & -3.7  & -2.8   & $(4 \pm 3) \times 10^{15}$ &    $6.12 \pm 0.02$  \\
$O_R(121)$ & -1.0  & no NSB  & lysogen monostable  & anti-imm.  unstable  \\
 \hline
$O_R(323)$ & -5.2  & -4.1   & $450 \pm 40$  & $450 \pm 30$   \\
$O_R(323)$ & -3.7  & -2.8   &  $(3 \pm 2) \times 10^{14}$ &   $190 \pm 20$  \\
$O_R(323)$ & -1.0  & no NSB  & $130 \pm 10$  &  $(2.5 \pm 0.6) \times 10^{10} $    \\
\hline
\end{tabular}
\end{center}
\caption{Spontaneous switching times (inverse of calculated switching rates)  for the Little mutants $O_R(121)$ and $O_R(323)$, computed as in Table 1 of the main text and assuming  a cell generation time of 34 min.} \label{tab:littlerates}
\end{table*}

We have used Forward Flux Sampling to compute spontaneous switching rates from the lysogenic to the anti-immune state, and {\em{vice versa}}, for the Little mutants $O_R(121)$ and $O_R(323)$. The computed  rates were converted into typical numbers of cell generations for which we expect these states to be stable, using a generation time of 34 min. The results are shown in \ref{tab:littlerates}, for a range of values of the parameters $\Delta G_\lp$ and $\Delta G_\nsb$.

We note that in modeling these mutants, one must decide which pair of CI dimers bound at $O_R$ experiences a cooperative binding interaction.  In our simulations, we assumed cooperative binding interactions for the first two adjacent CI dimers to bind.

\subsection{Effects of RNA polymerase concentration}\label{sec:Rp}

\begin{figure}[!h]
\includegraphics[width=.6\textwidth,clip=true]{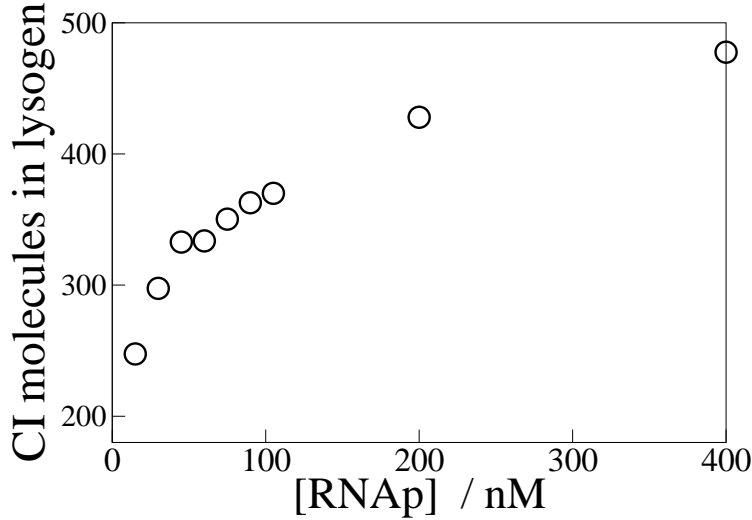}
\caption{ Effect of changing the free RNA polymerase concentration. The total number of CI molecules in the lysogenic state is plotted as a function of the free RNAp concentration. \label{fig:S3}}
\end{figure}

We tested the effect of the free RNA polymerase concentration, which was fixed at $50\nM$ in our model. This value was obtained by McClure, by fitting {\em{in vivo}} promoter expression data \cite{mcclure1983_proc}, but has recently been challenged \cite{Bremer2003}. We varied the free RNAp concentration and measured the total number of CI molecules in the lysogenic state. The results are plotted in Figure \ref{fig:S3}. Changing the free RNAp concentration can change the lysogenic CI concentration by about a factor of two. Although this parameter should be investigated further in future work, its value (within a reasonable physiological range) does not affect the qualitative features of our results. In fact, a higher free RNAp concentration is likely to further increase the stability of the lysogenic state. 

\section{Sensitivity of our conclusions to parameter choice}\label{sec:params}
\begin{figure}[!h]
\includegraphics[width=.6\textwidth,clip=true]{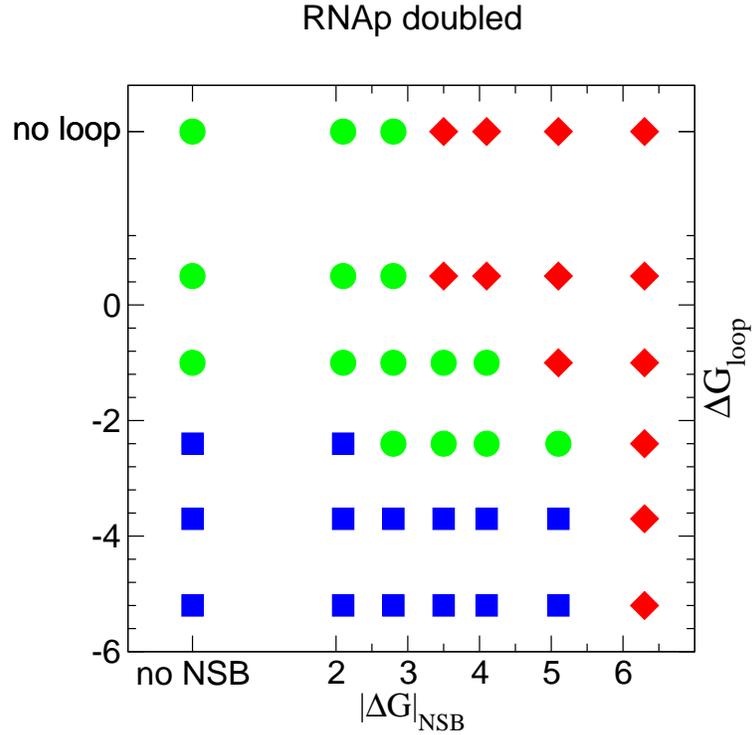}
\caption{ Range of bistability of the model switch as a function of the nonspecific DNA binding and DNA looping strengths, for a parameter set where the free RNAp concentration has been doubled [100nM instead of 50nM]. Comparing these results to those of Fig. 3 in the main text shows that our main conclusion, namely that DNA looping stabilises the switch against nonspecific DNA binding, also holds for these modified parameter sets. 
\label{fig:S4a}}
\end{figure}

\begin{figure}[!h]
\includegraphics[width=.6\textwidth,clip=true]{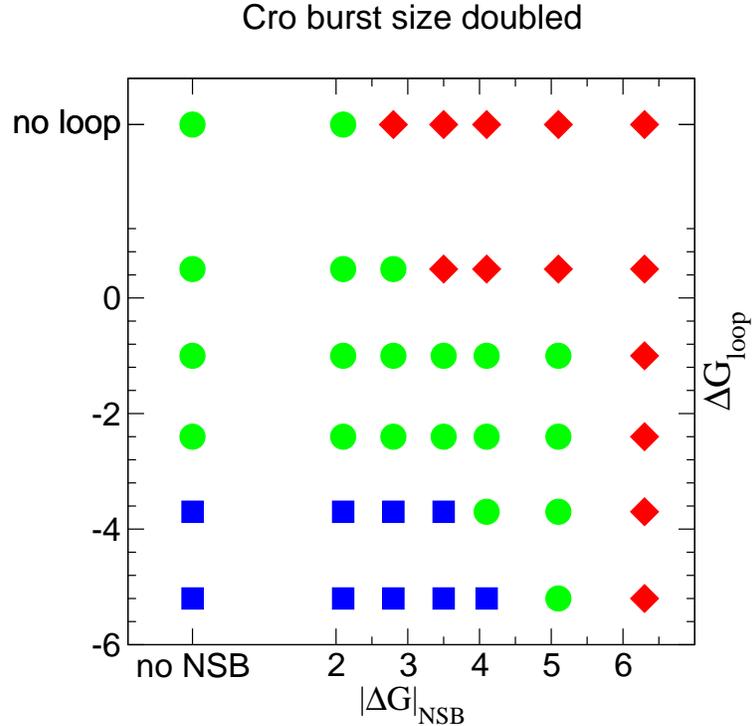}
\caption{Range of bistability of the model switch as a function of the nonspecific DNA binding and DNA looping strengths, for a parameter set where the average number of Cro molecules produced per mRNA transcript has been doubled [40 instead of 20] by doubling the translation rate while halving the transcription rate. Comparing these results to those of Fig. \ref{fig:phdiag} in the main text shows that our main conclusion, namely that DNA looping stabilises the switch against nonspecific DNA binding, also holds for these modified parameter sets.  \label{fig:S4b}}
\end{figure}

Our computational model involves several hundred chemical reactions. Many of the rate constants for these reactions are identical, so that the number of parameters is significantly less than the number of reactions. Nevertheless, parameter choice remains an important issue. We have taken most of the parameters for our model from the extensive biochemical data on the phage $\lambda$ switch available in the literature, as discussed in the main text. However, these values are subject to uncertainty, and in a few cases were unavailable, forcing us to make educated guesses (as detailed in the main text). It is therefore appropriate to investigate how sensitive our conclusions are to the choice of parameters. It is clearly impossible for us to prove that our conclusions hold throughout the many-dimensional space of all possible parameter variations. Nevertheless, we can strengthen our conclusions by testing the effects of several parameter changes, selected with the biology of the system in mind. To this end, we chose two parameters which are particularly uncertain, and are especially likely to influence the switching rate. We repeated our simulations for different values of these parameters. These are (1) the number of free RNA polymerase molecules in the cell, and (2)  the average number of Cro molecules produced per mRNA molecule (translational burst size). The free RNAp concentration has not been measured directly in experiments, and inferred values are subject to a large range of uncertainty. The translational burst size is also rather uncertain from biochemical data. The burst size for Cro is likely to be particularly important because is possible that a burst of Cro molecules produced from a single mRNA might be sufficient to flip the switch. We varied the burst size by increasing the translation rate while proportionately decreasing the transcription rate, so as to keep the average Cro level in the cell approximately constant. Figures \ref{fig:S4a} and \ref{fig:S4b} show the range of bistability for the model switch, as a function of the nonspecific DNA binding and DNA looping strengths, for free RNAp concentration doubled compared to the ``standard'' model [100nM instead of 50nM] \ref{fig:S4a} and Cro translational burst size doubled [40 per transcript instead of 20] \ref{fig:S4b}. Both these plots show the same qualitative behaviour as we obtained for the ``standard'' parameter set, as shown in Fig. \ref{fig:phdiag} of the main text. As the DNA looping strength increases, the range of nonspecific binding strengths over which the switch is bistable increases. We also used Forward Flux Sampling to compute switching rates for the lysogenic to anti-immune transition, in the case where the translational burst size for Cro was doubled. The results are shown in Table \ref{tab:burstrates}.

\begin{table*}[h]
\begin{center}
\begin{tabular}{|c|c|c|}
\hline
 $\Delta G_\lp$    & $\Delta G_\nsb$ & Switching time   \\
 $\kcm$ &  $\kcm$ & lys. $\to$ anti-imm.  \\
&  & (generations)  \\
\hline
no loop & no NSB & $(1.5 \pm 0.2) \times 10^{7}$  \\
no loop  & -2.1   & $(2.3 \pm 0.1) \times 10^{5}$  \\
-2.4  & no NSB   & $(1.5 \pm 0.4) \times 10^{11}$   \\
-2.4  & -2.1  & $(2.7 \pm 0.2) \times 10^{13}$   \\
\hline
\end{tabular}
\end{center}
\caption{Spontaneous switching times (inverse of calculated switching rates) for the model switch with Cro translational burst size doubled, by doubling the translation rate while halving the transcription rate. } \label{tab:burstrates}
\end{table*}          
Our main conclusions, that the lysogen is destabilized by nonspecific DNA binding but stabilized by the DNA looping interaction, remain the same with this modified parameter set. Comparing the two results without DNA looping in Table \ref{tab:burstrates}, the switching time in the presence of nonspecific DNA binding is two orders of magnitude faster than when nonspecific binding is absent. Comparing the results with and without DNA looping, we find that the looping interaction stabilises the lysogen by a factor of $10^4$ in the absence of nonspecific binding and by a factor of $10^6$ when nonspecific binding is present. These switching rates, together with the bistability analysis of Figs. \ref{fig:S4a} and \ref{fig:S4b}, suggest that the conclusions arising from this work are likely to be robust to uncertainties in the parameter values.

\section{Components and Reactions}\label{sec:lists}

\subsection{List of Components}

This is the list of components for our basic model, without non-specific binding and without looping. With the notation O(XYZ) we
refer to the operator O$_{\rm R}$ where X is bound to the site O$_{\rm R3}$, Y is bound to O$_{\rm R2}$ and Z is bound to O$_{\rm R1}$. 
$\{{\rm X,Y,Z}\}=\{{\rm 0,R,C,R_p}\}$, where 0 corresponds to an empty site, R corresponds to CI dimer bound, C to a Cro dimer bound, and ${\rm R_p}$
to an RNA polymerase molecule bound. The Cro promoter, $P_R$ overlaps with O$_{\rm R3}$, whereas the CI promoter, $P_{RM}$ overlaps with both 
O$_{\rm R2}$ and O$_{\rm R1}$, in our notation, RNA polymerase binds either to the X slot or to Y and Z simultaneously.

\noindent1	CI\\
2	Cro\\
3	CI$_2$\\
4	Cro$_2$\\
5	Rp\\
6	O(000)\\
7	O(R00)\\
8	O(0R0)\\
9	O(00R)\\
10	O(C00)\\
11	O(0C0)\\
12 O(00C)\\
13	O(Rp00)\\
14	O(0Rp)\\
15	O(RR0)\\
16	O(ROR)\\
17	O(RC0)\\
18	O(R0C)\\
19	O(RRp)\\
20	O(0RR)\\
21	O(CR0)\\
22	O(0RC)\\
23	O(RpR0\\
24	O(C0R)\\
25	O(OCR)\\
26	O(Rp0R)\\
27	O(CC0)\\
28	O(C0C)\\
29	O(CRp)\\
30	O(0CC)\\
31	O(RpC0)\\
32	O(Rp0C)\\
33	O(RRR)\\
34	O(RRC)\\
35	O(RCR)\\
36	O(CRR)\\
37	O(RpRR)\\
38	O(RCC)\\
39	O(CRC)\\
40	O(CCR)\\
41	O(RpCR)\\
42	O(RpRC)\\
43	O(CCC)\\
44	O(RpCC)\\
45	O(RpRp)\\
46	MCI\\
47	MCro\\
Adding nonspecific DNA binding requires 3 extra species:\\
48 D\\
49 DCI$_2$\\
50 DCro$_2$\\
When DNA looping is to be included in the model, we consider also the left operator O$_{\rm L}$, which also has three binding sites for transcription factors. RNA polymerase can bind to promoter $P_L$, which overlaps with the binding site O$_{\rm L1}$.  
The two operator can interact via a DNA loop stabilised by the octamerization of two CI tetramers which are already bound on
adjacent sites on the two operators. We label these looped states according to which pairs of CI dimers form the octamer. Looped states obtained by interaction between CI tetramers bound to O$_{\rm R1}$, O$_{\rm
R2}$,O$_{\rm L1}$ and O$_{\rm L2}$ are denoted OLR1.  States OLR2 have loops formed by interactions between tetramers bound to O$_{\rm R1}$, O$_{\rm R2}$, O$_{\rm L2}$ and O$_{\rm L3}$. 
States denoted  OLR3 have loops formed by interaction between tetramers bound to O$_{\rm R2}$, O$_{\rm R3}$, O$_{\rm L1}$ and O$_{\rm L2}$. 
Finally, states denoted OLR4 have loops formed by interaction between tetramers bound to O$_{\rm R2}$, O$_{\rm R3}$, O$_{\rm L2}$ and O$_{\rm L3}$. 
Each of these loop categories has two additional sites to which transcription factor dimers can bind. We therefore denote a particular looped configuration by OLRX(YZ), where X refers to loop type 1-4 (as above), Y denotes which species occupies the non-octamerized site on  O$_{\rm L}$ and Z labels the occupation of the non-octamerized site on O$_{\rm R}$. The extra species required in the model with looping are:\\
51	OL(000)\\
52	OL(R00)\\
53	OL(0R0)\\ 
54	OL(00R)\\
55	OL(C00)\\
56	OL(0C0)\\
57	OL(00C)\\
58	OL(00Rp)\\
59	OL(RR0)\\
60	OL(R0R)\\
61	OL(RCO)\\
62	OL(R0C)\\
63	OL(R0Rp)\\
64	OL(0RR)\\
65	OL(CR0)\\
66	OL(0RC)\\
67	OL(0RRp)\\
68	OL(C0R)\\
69	OL(0CR)\\
70	OL(CC0)\\
71	OL(C0C)\\
72	OL(C0Rp)\\
73	OL(0CC)\\
74	OL(0CRp)\\
75	OL(RRR)\\
76	OL(RRC)\\
77	OL(RRRp)\\
78	OL(RCR)\\
79	OL(CRR)\\
80	OL(RCC)\\
81	OL(RCRp)\\
82	OL(CRC)\\
83	OL(CRRp)\\
84	OL(CCR)\\
85	OL(CCC)\\
86	OL(CCRp)\\
87	OLR1(00)\\
88	OLR1(0R)\\
89	OLR1(0C)\\
90	OLR1(0Rp)\\
91	OLR1(R0)\\
92	OLR1(RR)\\
93	OLR1(RC)\\
94	OLR1(RRp)\\
95	OLR1(C0)\\
96	OLR1(CR)\\
97	OLR1(CC)\\
98	OLR1(CRp)\\
99	OLR2(00)\\
100	OLR2(0R)\\
101	OLR2(0C)\\
102	OLR2(0Rp)\\
103	OLR2(R0)\\
104 	OLR2(RR)\\
105	OLR2(RC)\\
106	OLR2(RRp)\\
107	OLR2(C0)\\
108	OLR2(CR)\\
109	OLR2(CC)\\
110	OLR2(CRp)\\
111	OLR3(00)\\
112	OLR3(0R)\\
113	OLR3(0C)\\
114	OLR3(R0)\\
115	OLR3(RR)\\
116	OLR3(RC)\\
117	OLR3(C0)\\
118	OLR3(CR)\\
119	OLR3(CC)\\
120	OLR4(00)\\
121	OLR4(0R)\\
122	OLR4(0C)\\
123	OLR4(R0)\\
124	OLR4(RR)\\
125	OLR4(RC)\\
126	OLR4(C0)\\
127	OLR4(CR)\\
128	OLR4(CC)\\
129	OLR2(Rp0)\\
130	OLR2(RpR)\\
131	OLR2(RpC)\\
132	OLR2(RpRp)\\
133	OLR4(Rp0)\\
134	OLR4(RpR)\\
135	OLR4(RpC)\\

\subsection{List of reactions}
\noindent This set is obtained with $\Delta G_{\rm loop}=-5.2$ kcal/mol and $\Delta G_{\rm NSB}=-3.5$ kcal/mol.\\
1) CI + CI  $\to$   CI$_2$ 		k = 0.628319\\
2) CI$_2$  $\to$   CI +  CI 		k = 5.705413\\
3) Cro + Cro  $\to$   Cro$_2$ 		k = 0.628319\\
4) Cro$_2$  $\to$   Cro +  Cro 		k = 280.199086\\
5) O(000) + CI$_2$  $\to$   O(R00) 		k = 0.314159\\
6) O(R00  $\to$   O(000) +  CI$_2$ 		k = 38.256936\\
7) O(000) + CI$_2$  $\to$   O(0R0) 		k = 0.314159\\
8) O(0R0)  $\to$   O(000) +  CI$_2$ 		k = 7.551827\\
9) O(000) + CI$_2$  $\to$   O(00R) 		k = 0.314159\\
10) O(00R)  $\to$   O(000) +  CI$_2$ 		k = 0.294263\\
11) O(000) + Cro$_2$  $\to$   O(C00) 		k = 0.314159\\
12) O(C00)  $\to$   O(000) +  Cro$_2$ 		k = 0.068319\\
13) O(000) + Cro$_2$  $\to$   O(0C0) 		k = 0.314159\\
14) O(0C0) $\to$   O(000) +  Cro$_2$ 		k = 4.641460\\
15) O(000) + Cro$_2$  $\to$   O(00C) 		k = 0.314159\\
16) O(00C)  $\to$   O(000) +  Cro$_2$ 		k = 0.662315\\
17) O(000) + Rp  $\to$   O(Rp00) 		k = 0.314159\\
18) O(Rp00)  $\to$   O(000) +  Rp 		k = 1.490712\\
19) O(000) + Rp  $\to$   O(0Rp) 		k = 0.314159\\
20) O(0Rp)  $\to$   O(000) +  Rp 		k = 0.294263\\
21) O(R00) + CI$_2$  $\to$   O(RR0) 		k = 0.314159\\
22) O(RR0)  $\to$   O(R00) +  CI$_2$ 		k = 0.068319\\
23) O(R00) + CI$_2$  $\to$   O(ROR) 		k = 0.314159\\
24) O(ROR)  $\to$   O(R00) +  CI$_2$ 		k = 0.294263\\
25) O(R00) + Cro$_2$  $\to$   O(RC0) 		k = 0.314159\\
26) O(RC0)  $\to$   O(R00) +  Cro$_2$ 		k = 4.641460\\
27) O(R00) + Cro$_2$  $\to$   O(R0C) 		k = 0.314159\\
28) O(R0C)  $\to$   O(R00) +  Cro$_2$ 		k = 0.662315\\
29) O(R00) + Rp  $\to$   O(RRp) 		k = 0.314159\\
30) O(RRp)  $\to$   O(R00) +  Rp 		k = 0.294263\\
31) O(0R0) + CI$_2$  $\to$   O(RR0) 		k = 0.314159\\
32) O(RR0)  $\to$   O(0R0) +  CI$_2$ 		k = 0.346100\\
33) O(0R0) + CI$_2$  $\to$   O(0RR) 		k = 0.314159\\
34) O(0RR)  $\to$   O(0R0) +  CI$_2$ 		k = 0.003683\\
35) O(0R0) + Cro$_2$  $\to$   O(CR0) 		k = 0.314159\\
36) O(CR0)  $\to$   O(0R0) +  Cro$_2$ 		k = 0.068319\\
37) O(0R0) + Cro$_2$  $\to$   O(0RC) 		k = 0.314159\\
38) O(0RC)  $\to$   O(0R0) +  Cro$_2$ 		k = 0.662315\\
39) O(0R0) + Rp  $\to$   O(RpR0) 		k = 0.314159\\
40) O(RpR0)  $\to$   O(0R0) +  Rp 		k = 1.490712\\
41) O(00R) + CI$_2$  $\to$   O(ROR) 		k = 0.314159\\
42) O(ROR)  $\to$   O(00R) +  CI$_2$ 		k = 38.256936\\
43) O(00R) + CI$_2$  $\to$   O(0RR) 		k = 0.314159\\
44) O(0RR)  $\to$   O(00R) +  CI$_2$ 		k = 0.094509\\
45) O(00R) + Cro$_2$  $\to$   O(C0R) 		k = 0.314159\\
46) O(C0R)  $\to$   O(00R) +  Cro$_2$ 		k = 0.068319\\
47) O(00R) + Cro$_2$  $\to$   O(0CR) 		k = 0.314159\\
48) O(0CR)  $\to$   O(00R) +  Cro$_2$ 		k = 4.641460\\
49) O(00R) + Rp  $\to$   O(Rp0R) 		k = 0.314159\\
50) O(Rp0R)  $\to$   O(00R) +  Rp 		k = 1.490712\\
51) O(C00) + CI$_2$  $\to$   O(CR0)		k = 0.314159\\
52) O(CR0)  $\to$   O(C00) +  CI$_2$ 		k = 7.551827\\
53) O(C00) + CI$_2$  $\to$   O(C0R) 		k = 0.314159\\
54) O(C0R)  $\to$   O(C00) +  CI$_2$ 		k = 0.294263\\
55) O(C00) + Cro$_2$  $\to$   O(CC0) 		k = 0.314159\\
56) O(CC0)  $\to$   O(C00) +  Cro$_2$ 		k = 1.753314\\
57) O(C00) + Cro$_2$  $\to$   O(C0C) 		k = 0.314159\\
58) O(C0C)  $\to$   O(C00) +  Cro$_2$ 		k = 0.662315\\
59) O(C00) + Rp  $\to$   O(CRp) 		k = 0.314159\\
60) O(CRp)  $\to$   O(C00) +  Rp 		k = 0.294263\\
61) O(0C0) + CI$_2$  $\to$   O(RC0) 		k = 0.314159\\
62) O(RC0)  $\to$   O(0C0) +  CI$_2$ 		k = 38.256936\\
63) O(0C0) + CI$_2$  $\to$   O(0CR) 		k = 0.314159\\
64) O(0CR)  $\to$   O(0C0) +  CI$_2$ 		k = 0.294263\\
65) O(0C0) + Cro$_2$  $\to$   O(CC0) 		k = 0.314159\\
66) O(CC0)  $\to$   O(0C0) +  Cro$_2$ 		k = 0.025808\\
67) O(0C0) + Cro$_2$  $\to$   O(0CC) 		k = 0.314159\\
68) O(0CC)  $\to$   O(0C0) +  Cro$_2$ 		k = 0.130739\\
69) O(0C0) + Rp  $\to$   O(RpC0) 		k = 0.314159\\
70) O(RpC0)  $\to$   O(0C0) +  Rp 		k = 1.490712\\
71) O(00C) + CI$_2$  $\to$   O(R0C) 		k = 0.314159\\
72) O(R0C)  $\to$   O(00C) +  CI$_2$ 		k = 38.256936\\
73) O(00C) + CI$_2$  $\to$   O(0RC) 		k = 0.314159\\
74) O(0RC)  $\to$   O(00C) +  CI$_2$ 		k = 7.551827\\
75) O(00C) + Cro$_2$  $\to$   O(C0C) 		k = 0.314159\\
76) O(C0C)  $\to$   O(00C) +  Cro$_2$ 		k = 0.068319\\
77) O(00C) + Cro$_2$  $\to$   O(0CC) 		k = 0.314159\\
78) O(0CC)  $\to$   O(00C) +  Cro$_2$ 		k = 0.916213\\
79) O(00C) + Rp  $\to$   O(Rp0C) 		k = 0.314159\\
80) O(Rp0C)  $\to$   O(00C) +  Rp 		k = 1.490712\\
81) O(Rp00) + CI$_2$  $\to$   O(RpR0) 		k = 0.314159\\
82) O(RpR0)  $\to$   O(Rp00) +  CI$_2$ 		k = 7.551827\\
83) O(Rp00) + CI$_2$  $\to$   O(Rp0R) 		k = 0.314159\\
84) O(Rp0R)  $\to$   O(Rp00) +  CI$_2$ 		k = 0.294263\\
85) O(Rp00) + Cro$_2$  $\to$   O(RpC0) 		k = 0.314159\\
86) O(RpC0)  $\to$   O(Rp00) +  Cro$_2$ 		k = 4.641460\\
87) O(Rp00) + Cro$_2$  $\to$   O(Rp0C) 		k = 0.314159\\
88) O(Rp0C)  $\to$   O(Rp00) +  Cro$_2$ 		k = 0.662315\\
89) O(Rp00) + Rp  $\to$   O(RpRp) 		k = 0.314159\\
90) O(RpRp)  $\to$   O(Rp00) +  Rp 		k = 0.294263\\
91) O(0Rp) + CI$_2$  $\to$   O(RRp) 		k = 0.314159\\
92) O(RRp)  $\to$   O(0Rp) +  CI$_2$ 		k = 38.256936\\
93) O(0Rp) + Cro$_2$  $\to$   O(CRp) 		k = 0.314159\\
94) O(CRp)  $\to$   O(0Rp) +  Cro$_2$ 		k = 0.068319\\
95) O(0Rp) + Rp  $\to$   O(RpRp) 		k = 0.314159\\
96) O(RpRp)  $\to$   O(0Rp) +  Rp 		k = 1.490712\\
97) O(RR0) + CI$_2$  $\to$   O(RRR) 		k = 0.314159\\
98) O(RRR)  $\to$   O(RR0) +  CI$_2$ 		k = 0.407068\\
99) O(RR0) + Cro$_2$  $\to$   O(RRC) 		k = 0.314159\\
100) O(RRC)  $\to$   O(RR0) +  Cro$_2$ 		k = 0.662315\\
101) O(ROR) + CI$_2$  $\to$   O(RRR) 		k = 0.314159\\
102) O(RRR)  $\to$   O(ROR) +  CI$_2$ 		k = 0.094509\\
103) O(ROR) + Cro$_2$  $\to$   O(RCR) 		k = 0.314159\\
104) O(RCR)  $\to$   O(ROR) +  Cro$_2$ 		k = 4.641460\\
105) O(0RR) + CI$_2$  $\to$   O(RRR) 		k = 0.314159\\
106) O(RRR)  $\to$   O(0RR) +  CI$_2$ 		k = 38.256936\\
107) O(0RR) + Cro$_2$  $\to$   O(CRR) 		k = 0.314159\\
108) O(CRR)  $\to$   O(0RR) +  Cro$_2$ 		k = 0.068319\\
109) O(0RR) + Rp  $\to$   O(RpRR) 		k = 0.314159\\
110) O(RpRR)  $\to$   O(0RR) +  Rp 		k = 1.490712\\
111) O(RC0) + CI$_2$  $\to$   O(RCR) 		k = 0.314159\\
112) O(RCR)  $\to$   O(RC0) +  CI$_2$ 		k = 0.294263\\
113) O(RC0) + Cro$_2$  $\to$   O(RCC) 		k = 0.314159\\
114) O(RCC)  $\to$   O(RC0) +  Cro$_2$ 		k = 0.130739\\
115) O(R0C) + CI$_2$  $\to$   O(RRC) 		k = 0.314159\\
116) O(RRC)  $\to$   O(R0C) +  CI$_2$ 		k = 0.068319\\
117) O(R0C) + Cro$_2$  $\to$   O(RCC) 		k = 0.314159\\
118) O(RCC) $\to$   O(R0C) +  Cro$_2$ 		k = 1.753314\\
119) O(CR0) + CI$_2$  $\to$   O(CRR) 		k = 0.314159\\
120) O(CRR)  $\to$   O(CR0) +  CI$_2$ 		k = 0.003683\\
121) O(CR0) + Cro$_2$  $\to$   O(CRC) 		k = 0.314159\\
122) O(CRC)  $\to$   O(CR0) +  Cro$_2$ 		k = 0.662315\\
123) O(C0R) + CI$_2$  $\to$   O(CRR) 		k = 0.314159\\
124) O(CRR)  $\to$   O(C0R) +  CI$_2$ 		k = 0.094509\\
125) O(C0R) + Cro$_2$  $\to$   O(CCR) 		k = 0.314159\\
126) O(CCR)  $\to$   O(C0R) +  Cro$_2$ 		k = 1.753314\\
127) O(0CR) + CI$_2$  $\to$   O(RCR) 		k = 0.314159\\
128) O(RCR)  $\to$   O(0CR) +  CI$_2$ 		k = 38.256936\\
129) O(0CR) + Cro$_2$  $\to$   O(CCR) 		k = 0.314159\\
130) O(CCR)  $\to$   O(0CR) +  Cro$_2$ 		k = 0.025808\\
131) O(0CR) + Rp  $\to$   O(RpCR) 		k = 0.314159\\
132) O(RpCR)  $\to$   O(0CR) +  Rp 		k = 1.490712\\
133) O(0RC) + CI$_2$  $\to$   O(RRC) 		k = 0.314159\\
134) O(RRC)  $\to$   O(0RC) +  CI$_2$ 		k = 0.346100\\
135) O(0RC) + Cro$_2$  $\to$   O(CRC) 		k = 0.314159\\
136) O(CRC)  $\to$   O(0RC) +  Cro$_2$ 		k = 0.068319\\
137) O(0RC) + Rp  $\to$   O(RpRC) 		k = 0.314159\\
138) O(RpRC)  $\to$   O(0RC) +  Rp 		k = 1.490712\\
139) O(CC0) + CI$_2$  $\to$   O(CCR) 		k = 0.314159\\
140) O(CCR)  $\to$   O(CC0) +  CI$_2$ 		k = 0.294263\\
141) O(CC0) + Cro$_2$  $\to$   O(CCC) 		k = 0.314159\\
142) O(CCC)  $\to$   O(CC0) +  Cro$_2$ 		k = 0.407068\\
143) O(C0C) + CI$_2$  $\to$   O(CRC) 		k = 0.314159\\
144) O(CRC)  $\to$   O(C0C) +  CI$_2$ 		k = 7.551827\\
145) O(C0C) + Cro$_2$  $\to$   O(CCC) 		k = 0.314159\\
146) O(CCC) $\to$   O(C0C) +  Cro$_2$ 		k = 1.077612\\
147) O(0CC) + CI$_2$  $\to$   O(RCC) 		k = 0.314159\\
148) O(RCC)  $\to$   O(0CC) +  CI$_2$ 		k = 38.256936\\
149) O(0CC) + Cro$_2$  $\to$   O(CCC) 		k = 0.314159\\
150) O(CCC)  $\to$   O(0CC) +  Cro$_2$ 		k = 0.080354\\
151) O(0CC) + Rp  $\to$   O(RpCC) 		k = 0.314159\\
152) O(RpCC)  $\to$   O(0CC) +  Rp 		k = 1.490712\\
153) O(RpR0) + CI$_2$  $\to$   O(RpRR) 		k = 0.314159\\
154) O(RpRR)  $\to$   O(RpR0) +  CI$_2$ 		k = 0.003683\\
155) O(RpR0) + Cro$_2$  $\to$   O(RpRC) 		k = 0.314159\\
156) O(RpRC)  $\to$   O(RpR0) +  Cro$_2$ 		k = 0.662315\\
157) O(Rp0R) + CI$_2$  $\to$   O(RpRR) 		k = 0.314159\\
158) O(RpRR)  $\to$   O(Rp0R) +  CI$_2$ 		k = 0.094509\\
159) O(Rp0R) + Cro$_2$  $\to$   O(RpCR) 		k = 0.314159\\
160) O(RpCR)  $\to$   O(Rp0R) +  Cro$_2$ 		k = 4.641460\\
161) O(RpC0) + CI$_2$  $\to$   O(RpCR) 		k = 0.314159\\
162) O(RpCR)  $\to$   O(RpC0) +  CI$_2$ 		k = 0.294263\\
163) O(RpC0) + Cro$_2$  $\to$   O(RpCC) 		k = 0.314159\\
164) O(RpCC)  $\to$   O(RpC0) +  Cro$_2$ 		k = 0.130739\\
165) O(Rp0C) + CI$_2$  $\to$   O(RpRC) 		k = 0.314159\\
166) O(RpRC)  $\to$   O(Rp0C) +  CI$_2$ 		k = 7.551827\\
167) O(Rp0C) + Cro$_2$  $\to$   O(RpCC) 		k = 0.314159\\
168) O(RpCC)  $\to$   O(Rp0C) +  Cro$_2$ 		k = 1.753314\\
169) O(Rp00)  $\to$   O(000) +  Rp +  MCI 		k = 0.001000\\
170) O(RpR0)  $\to$   O(0R0) +  Rp +  MCI 		k = 0.011000\\
171) O(Rp0R)  $\to$   O(00R) +  Rp +  MCI 		k = 0.001000\\
172) O(RpC0)  $\to$   O(0C0) +  Rp +  MCI 		k = 0.001000\\
173) O(Rp0C)  $\to$   O(00C) +  Rp +  MCI 		k = 0.001000\\
174) O(RpRR)  $\to$   O(0RR) +  Rp +  MCI 		k = 0.011000\\
175) O(RpRC)  $\to$   O(0RC) +  Rp +  MCI 		k = 0.011000\\
176) O(RpCR)  $\to$   O(0CR) +  Rp +  MCI 		k = 0.001000\\
177) O(RpCC)  $\to$   O(0CC) +  Rp +  MCI 		k = 0.001000\\
178) O(RpRp)  $\to$   O(0Rp) +  Rp +  MCI 		k = 0.001000\\
179) O(0Rp) $\to$   O(000)+  Rp +  MCro 		k = 0.014000\\
180) O(RRp)  $\to$   O(R00) +  Rp +  MCro 		k = 0.014000\\
181) O(CRp)  $\to$   O(C00) +  Rp +  MCro 		k = 0.014000\\
182) O(RpRp)  $\to$   O(Rp00) +  Rp +  MCro 		k = 0.014000\\
183) MCI  $\to$   MCI +  CI 		k = 0.034656\\
184) MCro  $\to$   MCro +  Cro 		k = 0.115520\\
185) MCI  $\to$  0 		k = 0.005776\\
186) MCro  $\to$  0 		k = 0.005776\\
187) CI  $\to$  0 		k = 0.000340\\
188) Cro  $\to$  0 		k = 0.000606\\
189) CI$_2$  $\to$  0 		k = 0.000340\\
190) Cro$_2$  $\to$  0 		k = 0.000340\\

\noindent If the system includes looping, the reaction schemes should be extended with the following reactions (rate constants calculated for  $\Delta G_{\rm loop}=-5.2$ kcal/mol):

\noindent 191) OL(000) + CI$_2$  $\to$   OL(R00) 		k = 0.314159\\
192) OL(R00)  $\to$   OL(000) +  CI$_2$ 		k = 0.346100\\
193) OL(000) + CI$_2$  $\to$   OL(0R0) 		k = 0.314159\\
194) OL(0R0)  $\to$   OL(000) +  CI$_2$ 		k = 0.563117\\
195) OL(000) + CI$_2$  $\to$   OL(00R) 		k = 0.314159\\
196) OL(00R)  $\to$   OL(000) +  CI$_2$ 		k = 0.035701\\
197) OL(000) + Cro$_2$  $\to$   OL(C00) 		k = 0.314159\\
198) OL(C00)  $\to$   OL(000) +  Cro$_2$ 		k = 0.068319\\
199) OL(000) + Cro$_2$  $\to$   OL(0C0) 		k = 0.314159\\
200) OL(0C0)  $\to$   OL(000) +  Cro$_2$ 		k = 4.641460\\
201) OL(000) + Cro$_2$  $\to$   OL(00C) 		k = 0.314159\\
202) OL(00C)  $\to$   OL(000) +  Cro$_2$ 		k = 0.662315\\
203) OL(000) + Rp  $\to$   OL(00Rp) 		k = 0.314159\\
204) OL(00Rp)  $\to$   OL(000) +  Rp 		k = 2.062175\\
205) OL(R00) + CI$_2$  $\to$   OL(RR0) 		k = 0.314159\\
206) OL(RR0)  $\to$   OL(R00) +  CI$_2$ 		k = 0.009749\\
207) OL(R00) + CI$_2$  $\to$   OL(R0R) 		k = 0.314159\\
208) OL(R0R)  $\to$   OL(R00) +  CI$_2$ 		k = 0.035701\\
209) OL(R00) + Cro$_2$  $\to$   OL(RCO) 		k = 0.314159\\
210) OL(RCO)  $\to$   OL(R00) +  Cro$_2$ 		k = 4.641460\\
211) OL(R00) + Cro$_2$  $\to$   OL(R0C) 		k = 0.314159\\
212) OL(R0C)  $\to$   OL(R00) +  Cro$_2$ 		k = 0.662315\\
213) OL(R00) + Rp  $\to$   OL(R0Rp) 		k = 0.314159\\
214) OL(R0Rp)  $\to$   OL(R00) +  Rp 		k = 2.062175\\
215) OL(0R0) + CI$_2$  $\to$   OL(RR0 		k = 0.314159\\
216) OL(RR0)  $\to$   OL(0R0) +  CI$_2$ 		k = 0.005992\\
217) OL(0R0) + CI$_2$  $\to$   OL(0RR) 		k = 0.314159\\
218) OL(0RR)  $\to$   OL(0R0) +  CI$_2$ 		k = 0.000618\\
219) OL(0R0) + Cro$_2$  $\to$   OL(CR0) 		k = 0.314159\\
220) OL(CR0)  $\to$   OL(0R0) +  Cro$_2$ 		k = 0.068319\\
221) OL(0R0) + Cro$_2$  $\to$   OL(0RC) 		k = 0.314159\\
222) OL(0RC)  $\to$   OL(0R0) +  Cro$_2$ 		k = 0.662315\\
223) OL(0R0) + Rp  $\to$   OL(0RRp) 		k = 0.314159\\
224) OL(0RRp)  $\to$   OL(0R0) +  Rp 		k = 2.062175\\
225) OL(00R) + CI$_2$  $\to$   OL(R0R) 		k = 0.314159\\
226) OL(R0R)  $\to$   OL(00R) +  CI$_2$ 		k = 0.346100\\
227) OL(00R) + CI$_2$  $\to$   OL(0RR) 		k = 0.314159\\
228) OL(0RR)  $\to$   OL(00R) +  CI$_2$ 		k = 0.009749\\
229) OL(00R) + Cro$_2$  $\to$   OL(C0R) 		k = 0.314159\\
230) OL(C0R)  $\to$   OL(00R) +  Cro$_2$ 		k = 0.068319\\
231) OL(00R) + Cro$_2$  $\to$   OL(0CR)		k = 0.314159\\
232) OL(0CR)  $\to$   OL(00R) +  Cro$_2$ 		k = 4.641460\\
233) OL(C00) + CI$_2$  $\to$   OL(CR0) 		k = 0.314159\\
234) OL(CR0)  $\to$   OL(C00) +  CI$_2$ 		k = 0.563117\\
235) OL(C00) + CI$_2$  $\to$   OL(C0R) 		k = 0.314159\\
236) OL(C0R)  $\to$   OL(C00) +  CI$_2$ 		k = 0.035701\\
237) OL(C00) + Cro$_2$  $\to$   OL(CC0) 		k = 0.314159\\
238) OL(CC0)  $\to$   OL(C00) +  Cro$_2$ 		k = 1.753314\\
239) OL(C00) + Cro$_2$  $\to$   OL(C0C) 		k = 0.314159\\
240) OL(C0C)  $\to$   OL(C00) +  Cro$_2$ 		k = 0.662315\\
241) OL(C00) + Rp  $\to$   OL(C0Rp) 		k = 0.314159\\
242) OL(C0Rp)  $\to$   OL(C00) +  Rp 		k = 2.062175\\
243) OL(0C0) + CI$_2$  $\to$   OL(RCO) 		k = 0.314159\\
244) OL(RCO)  $\to$   OL(0C0) +  CI$_2$ 		k = 0.346100\\
245) OL(0C0) + CI$_2$  $\to$   OL(0CR) 		k = 0.314159\\
246) OL(0CR)  $\to$   OL(0C0) +  CI$_2$ 		k = 0.035701\\
247) OL(0C0) + Cro$_2$  $\to$   OL(CC0) 		k = 0.314159\\
248) OL(CC0)  $\to$   OL(0C0) +  Cro$_2$ 		k = 0.025808\\
249) OL(0C0) + Cro$_2$  $\to$   OL(0CC) 		k = 0.314159\\
250) OL(0CC)  $\to$   OL(0C0) +  Cro$_2$ 		k = 0.130739\\
251) OL(0C0) + Rp  $\to$   OL(0CRp) 		k = 0.314159\\
252) OL(0CRp)  $\to$   OL(0C0) +  Rp 		k = 2.062175\\
253) OL(00C) + CI$_2$  $\to$   OL(R0C) 		k = 0.314159\\
254) OL(R0C)  $\to$   OL(00C) +  CI$_2$ 		k = 0.346100\\
255) OL(00C) + CI$_2$  $\to$   OL(0RC) 		k = 0.314159\\
256) OL(0RC)  $\to$   OL(00C) +  CI$_2$ 		k = 0.563117\\
257) OL(00C) + Cro$_2$  $\to$   OL(C0C) 		k = 0.314159\\
258) OL(C0C)  $\to$   OL(00C) +  Cro$_2$ 		k = 0.068319\\
259) OL(00C) + Cro$_2$  $\to$   OL(0CC) 		k = 0.314159\\
260) OL(0CC)  $\to$   OL(00C) +  Cro$_2$ 		k = 0.916213\\
261) OL(00Rp) + CI$_2$  $\to$   OL(R0Rp) 		k = 0.314159\\
262) OL(R0Rp)  $\to$   OL(00Rp) +  CI$_2$ 		k = 0.346100\\
263) OL(00Rp) + CI$_2$  $\to$   OL(0RRp) 		k = 0.314159\\
264) OL(0RRp)  $\to$   OL(00Rp) +  CI$_2$ 		k = 0.563117\\
265) OL(00Rp) + Cro$_2$  $\to$   OL(C0Rp) 		k = 0.314159\\
266) OL(C0Rp)  $\to$   OL(00Rp) +  Cro$_2$ 		k = 0.068319\\
267) OL(00Rp) + Cro$_2$  $\to$   OL(0CRp) 		k = 0.314159\\
268) OL(0CRp)  $\to$   OL(00Rp) +  Cro$_2$ 		k = 4.641460\\
269) OL(RR0) + CI$_2$  $\to$   OL(RRR) 		k = 0.314159\\
270) OL(RRR)  $\to$   OL(RR0) +  CI$_2$ 		k = 0.035701\\
271) OL(RR0) + Cro$_2$  $\to$   OL(RRC) 		k = 0.314159\\
272) OL(RRC)  $\to$   OL(RR0) +  Cro$_2$ 		k = 0.662315\\
273) OL(RR0) + Rp  $\to$   OL(RRRp) 		k = 0.314159\\
274) OL(RRRp)  $\to$   OL(RR0) +  Rp 		k = 2.062175\\
275) OL(R0R) + CI$_2$  $\to$   OL(RRR) 		k = 0.314159\\
276) OL(RRR)  $\to$   OL(R0R) +  CI$_2$ 		k = 0.009749\\
277) OL(R0R) + Cro$_2$  $\to$   OL(RCR) 		k = 0.314159\\
278) OL(RCR)  $\to$   OL(R0R) +  Cro$_2$ 		k = 4.641460\\
279) OL(0RR) + CI$_2$  $\to$   OL(RRR) 		k = 0.314159\\
280) OL(RRR)  $\to$   OL(0RR) +  CI$_2$ 		k = 0.346100\\
281) OL(0RR) + Cro$_2$  $\to$   OL(CRR) 		k = 0.314159\\
282) OL(CRR)  $\to$   OL(0RR) +  Cro$_2$ 		k = 0.068319\\
283) OL(RCO) + CI$_2$  $\to$   OL(RCR) 		k = 0.314159\\
284) OL(RCR)  $\to$   OL(RCO) +  CI$_2$ 		k = 0.035701\\
285) OL(RCO) + Cro$_2$  $\to$   OL(RCC) 		k = 0.314159\\
286) OL(RCC)  $\to$   OL(RCO) +  Cro$_2$ 		k = 0.130739\\
287) OL(RCO) + Rp  $\to$   OL(RCRp) 		k = 0.314159\\
288) OL(RCRp)  $\to$   OL(RCO) +  Rp 		k = 2.062175\\
289) OL(R0C) + CI$_2$  $\to$   OL(RRC) 		k = 0.314159\\
290) OL(RRC)  $\to$   OL(R0C) +  CI$_2$ 		k = 0.009749\\
291) OL(R0C) + Cro$_2$  $\to$   OL(RCC) 		k = 0.314159\\
292) OL(RCC)  $\to$   OL(R0C) +  Cro$_2$ 		k = 1.753314\\
293) OL(CR0) + CI$_2$  $\to$   OL(CRR) 		k = 0.314159\\
294) OL(CRR)  $\to$   OL(CR0) +  CI$_2$ 		k = 0.000618\\
295) OL(CR0) + Cro$_2$  $\to$   OL(CRC) 		k = 0.314159\\
296) OL(CRC)  $\to$   OL(CR0) +  Cro$_2$ 		k = 0.662315\\
297) OL(CR0) + Rp  $\to$   OL(CRRp) 		k = 0.314159\\
298) OL(CRRp)  $\to$   OL(CR0) +  Rp 		k = 2.062175\\
299) OL(C0R) + CI$_2$  $\to$   OL(CRR) 		k = 0.314159\\
300) OL(CRR)  $\to$   OL(C0R) +  CI$_2$ 		k = 0.009749\\
301) OL(C0R) + Cro$_2$  $\to$   OL(CCR) 		k = 0.314159\\
302) OL(CCR)  $\to$   OL(C0R) +  Cro$_2$ 		k = 1.753314\\
303) OL(0CR) + CI$_2$  $\to$   OL(RCR) 		k = 0.314159\\
304) OL(RCR)  $\to$   OL(0CR) +  CI$_2$ 		k = 0.346100\\
305) OL(0CR) + Cro$_2$  $\to$   OL(CCR) 		k = 0.314159\\
306) OL(CCR)  $\to$   OL(0CR) +  Cro$_2$ 		k = 0.025808\\
307) OL(0RC) + CI$_2$  $\to$   OL(RRC) 		k = 0.314159\\
308) OL(RRC)  $\to$   OL(0RC) +  CI$_2$ 		k = 0.005992\\
309) OL(0RC) + Cro$_2$  $\to$   OL(CRC) 		k = 0.314159\\
310) OL(CRC)  $\to$   OL(0RC) +  Cro$_2$ 		k = 0.068319\\
311) OL(CC0) + CI$_2$  $\to$   OL(CCR) 		k = 0.314159\\
312) OL(CCR)  $\to$   OL(CC0) +  CI$_2$ 		k = 0.035701\\
313) OL(CC0) + Cro$_2$  $\to$   OL(CCC) 		k = 0.314159\\
314) OL(CCC)  $\to$   OL(CC0) +  Cro$_2$ 		k = 0.407068\\
315) OL(CC0) + Rp  $\to$   OL(CCRp) 		k = 0.314159\\
316) OL(CCRp)  $\to$   OL(CC0) +  Rp 		k = 2.062175\\
317) OL(C0C) + CI$_2$  $\to$   OL(CRC) 		k = 0.314159\\
318) OL(CRC)  $\to$   OL(C0C) +  CI$_2$ 		k = 0.563117\\
319) OL(C0C) + Cro$_2$  $\to$   OL(CCC) 		k = 0.314159\\
320) OL(CCC)  $\to$   OL(C0C) +  Cro$_2$ 		k = 1.077612\\
321) OL(0CC) + CI$_2$  $\to$   OL(RCC) 		k = 0.314159\\
322) OL(RCC)  $\to$   OL(0CC) +  CI$_2$ 		k = 0.346100\\
323) OL(0CC) + Cro$_2$  $\to$   OL(CCC) 		k = 0.314159\\
324) OL(CCC)  $\to$   OL(0CC) +  Cro$_2$ 		k = 0.080354\\
325) OL(0RRp) + CI$_2$  $\to$   OL(RRRp) 		k = 0.314159\\
326) OL(RRRp)  $\to$   OL(0RRp) +  CI$_2$ 		k = 0.005992\\
327) OL(0RRp) + Cro$_2$  $\to$   OL(CRRp) 		k = 0.314159\\
328) OL(CRRp)  $\to$   OL(0RRp) +  Cro$_2$ 		k = 0.068319\\
329) OL(R0Rp) + CI$_2$  $\to$   OL(RRRp) 		k = 0.314159\\
330) OL(RRRp)  $\to$   OL(R0Rp) +  CI$_2$ 		k = 0.009749\\
331) OL(R0Rp) + Cro$_2$  $\to$   OL(RCRp) 		k = 0.314159\\
332) OL(RCRp)  $\to$   OL(R0Rp) +  Cro$_2$ 		k = 4.641460\\
333) OL(0CRp) + CI$_2$  $\to$   OL(RCRp) 		k = 0.314159\\
334) OL(RCRp)  $\to$   OL(0CRp) +  CI$_2$ 		k = 0.346100\\
335) OL(0CRp) + Cro$_2$  $\to$   OL(CCRp) 		k = 0.314159\\
336) OL(CCRp)  $\to$   OL(0CRp) +  Cro$_2$ 		k = 0.025808\\
337) OL(C0Rp) + CI$_2$  $\to$   OL(CRRp) 		k = 0.314159\\
338) OL(CRRp)  $\to$   OL(C0Rp) +  CI$_2$ 		k = 0.563117\\
339) OL(C0Rp) + Cro$_2$  $\to$   OL(CCRp) 		k = 0.314159\\
340) OL(CCRp)  $\to$   OL(C0Rp) +  Cro$_2$ 		k = 1.753314\\
341) OL(0RR) + O(0RR)  $\to$   OLR1(00) 		k = 62.095095\\
342) OLR1(00)  $\to$   OL(0RR) +  O(0RR) 		k = 0.008046\\
343) OL(0RR) + O(RRR)  $\to$   OLR1(0R) 		k = 62.095095\\
344) OLR1(0R)  $\to$   OL(0RR) +  O(RRR) 		k = 0.008046\\
345) OL(0RR) + O(CRR)  $\to$   OLR1(0C) 		k = 62.095095\\
346) OLR1(0C)  $\to$   OL(0RR) +  O(CRR) 		k = 0.008046\\
347) OL(0RR) + O(RpRR)  $\to$   OLR1(0Rp) 		k = 62.095095\\
348) OLR1(0Rp)  $\to$   OL(0RR) +  O(RpRR) 		k = 0.008046\\
349) OL(RRR) + O(0RR)  $\to$   OLR1(R0) 		k = 62.095095\\
350) OLR1(R0)  $\to$   OL(RRR) +  O(0RR) 		k = 0.008046\\
351) OL(RRR) + O(RRR)  $\to$   OLR1(RR) 		k = 62.095095\\
352) OLR1(RR)  $\to$   OL(RRR) +  O(RRR) 		k = 0.008046\\
353) OL(RRR) + O(CRR)  $\to$   OLR1(RC) 		k = 62.095095\\
354) OLR1(RC)  $\to$   OL(RRR) +  O(CRR) 		k = 0.008046\\
355) OL(RRR) + O(RpRR)  $\to$   OLR1(RRp) 		k = 62.095095\\
356) OLR1(RRp)  $\to$   OL(RRR) +  O(RpRR) 		k = 0.008046\\
357) OL(CRR) + O(0RR)  $\to$   OLR1(C0) 		k = 62.095095\\
358) OLR1(C0)  $\to$   OL(CRR) +  O(0RR) 		k = 0.008046\\
359) OL(CRR) + O(RRR)  $\to$   OLR1(CR) 		k = 62.095095\\
360) OLR1(CR)  $\to$   OL(CRR) +  O(RRR) 		k = 0.008046\\
361) OL(CRR) + O(CRR)  $\to$   OLR1(CC) 		k = 62.095095\\
362) OLR1(CC)  $\to$   OL(CRR) +  O(CRR) 		k = 0.008046\\
363) OL(CRR) + O(RpRR)  $\to$   OLR1(CRp) 		k = 62.095095\\
364) OLR1(CRp)  $\to$   OL(CRR) +  O(RpRR) 		k = 0.008046\\
365) OLR1(00) + CI$_2$  $\to$   OLR1(R0) 		k = 0.314159\\
366) OLR1(R0)  $\to$   OLR1(00) +  CI$_2$ 		k = 0.346100\\
367) OLR1(00)+ CI$_2$  $\to$   OLR1(0R) 		k = 0.314159\\
368) OLR1(0R)  $\to$   OLR1(00) +  CI$_2$ 		k = 38.256936\\
369) OLR1(00) + Cro$_2$  $\to$   OLR1(C0) 		k = 0.314159\\
370) OLR1(C0)  $\to$   OLR1(00) +  Cro$_2$ 		k = 0.068319\\
371) OLR1(00) + Cro$_2$  $\to$   OLR1(0C) 		k = 0.314159\\
372) OLR1(0C)  $\to$   OLR1(00) +  Cro$_2$ 		k = 0.068319\\
373) OLR1(00) + Rp  $\to$   OLR1(0Rp)		k = 0.314159\\
374) OLR1(0Rp)  $\to$   OLR1(00) +  Rp 		k = 1.490712\\
375) OLR1(R0) + CI$_2$  $\to$   OLR1(RR) 		k = 0.314159\\
376) OLR1(RR)  $\to$   OLR1(R0) +  CI$_2$ 		k = 0.294263\\
377) OLR1(R0) + Cro$_2$  $\to$   OLR1(RC) 		k = 0.314159\\
378) OLR1(RC)  $\to$   OLR1(R0) +  Cro$_2$ 		k = 0.068319\\
379) OLR1(R0) + Rp  $\to$   OLR1(RRp) 		k = 0.314159\\
380) OLR1(RRp)  $\to$   OLR1(R0) +  Rp 		k = 1.490712\\
381) OLR1(C0) + CI$_2$  $\to$   OLR1(CR) 		k = 0.314159\\
382) OLR1(CR)  $\to$   OLR1(C0) +  CI$_2$ 		k = 38.256936\\
383) OLR1(C0) + Cro$_2$  $\to$   OLR1(CC) 		k = 0.314159\\
384) OLR1(CC)  $\to$   OLR1(C0) +  Cro$_2$ 		k = 0.068319\\
385) OLR1(C0) + Rp  $\to$   OLR1(CRp) 		k = 0.314159\\
386) OLR1(CRp)  $\to$   OLR1(C0) +  Rp 		k = 1.490712\\
387) OLR1(0R) + CI$_2$  $\to$   OLR1(RR) 		k = 0.314159\\
388) OLR1(RR)  $\to$   OLR1(0R) +  CI$_2$ 		k = 0.002662\\
389) OLR1(0R) + Cro$_2$  $\to$   OLR1(CR) 		k = 0.314159\\
390) OLR1(CR)  $\to$   OLR1(0R) +  Cro$_2$ 		k = 0.068319\\
391) OLR1(0C) + CI$_2$  $\to$   OLR1(RC) 		k = 0.314159\\
392) OLR1(RC)  $\to$   OLR1(0C) +  CI$_2$ 		k = 0.346100\\
393) OLR1(0C) + Cro$_2$  $\to$   OLR1(CC) 		k = 0.314159\\
394) OLR1(CC)  $\to$   OLR1(0C) +  Cro$_2$ 		k = 0.068319\\
395) OLR1(0Rp) + CI$_2$  $\to$   OLR1(RRp) 		k = 0.314159\\
396) OLR1(RRp)  $\to$   OLR1(0Rp)+  CI$_2$ 		k = 0.346100\\
397) OLR1(0Rp) + Cro$_2$  $\to$   OLR1(CRp) 		k = 0.314159\\
398) OLR1(CRp)  $\to$   OLR1(0Rp) +  Cro$_2$ 		k = 0.068319\\
399) OLR1(0Rp)  $\to$   OLR1(00) +  Rp +  MCI 		k = 0.011000\\
400) OLR1(RRp)  $\to$   OLR1(R0) +  Rp +  MCI 		k = 0.011000\\
401) OLR1(CRp)  $\to$   OLR1(C0) +  Rp +  MCI 		k = 0.011000\\
402) OL(RR0) + O(0RR)  $\to$   OLR2(00) 		k = 62.095095\\
403) OLR2(00)  $\to$   OL(RR0) +  O(0RR) 		k = 0.008046\\
404) OL(RR0) + O(RRR)  $\to$   OLR2(0R) 		k = 62.095095\\
405) OLR2(0R)  $\to$   OL(RR0) +  O(RRR) 		k = 0.008046\\
406) OL(RR0) + O(CRR)  $\to$   OLR2(0C) 		k = 62.095095\\
407) OLR2(0C)  $\to$   OL(RR0) +  O(CRR) 		k = 0.008046\\
408) OL(RR0) + O(RpRR)  $\to$   OLR2(0Rp) 		k = 62.095095\\
409) OLR2(0Rp)  $\to$   OL(RR0) +  O(RpRR) 		k = 0.008046\\
410) OL(RRR) + O(0RR)  $\to$   OLR2(R0) 		k = 62.095095\\
411) OLR2(R0)  $\to$   OL(RRR) +  O(0RR) 		k = 0.008046\\
412) OL(RRR) + O(RRR)  $\to$   OLR2(RR) 		k = 62.095095\\
413) OLR2(RR)  $\to$   OL(RRR) +  O(RRR) 		k = 0.008046\\
414) OL(RRR) + O(CRR)  $\to$   OLR2(RC) 		k = 62.095095\\
415) OLR2(RC)  $\to$   OL(RRR) +  O(CRR) 		k = 0.008046\\
416) OL(RRR) + O(RpRR)  $\to$   OLR2(RRp) 		k = 62.095095\\
417) OLR2(RRp)  $\to$   OL(RRR) +  O(RpRR) 		k = 0.008046\\
418) OL(RRC) + O(0RR)  $\to$   OLR2(C0) 		k = 62.095095\\
419) OLR2(C0)  $\to$   OL(RRC) +  O(0RR) 		k = 0.008046\\
420) OL(RRC) + O(RRR)  $\to$   OLR2(CR) 		k = 62.095095\\
421) OLR2(CR)  $\to$   OL(RRC) +  O(RRR) 		k = 0.008046\\
422) OL(RRC) + O(CRR)  $\to$   OLR2(CC) 		k = 62.095095\\
423) OLR2(CC)  $\to$   OL(RRC) +  O(CRR) 		k = 0.008046\\
424) OL(RRC) + O(RpRR)  $\to$   OLR2(CRp) 		k = 62.095095\\
425) OLR2(CRp)  $\to$   OL(RRC) +  O(RpRR) 		k = 0.008046\\
426) OL(RRRp) + O(0RR)  $\to$   OLR2(Rp0) 		k = 62.095095\\
427) OLR2(Rp0)  $\to$   OL(RRRp) +  O(0RR) 		k = 0.008046\\
428) OL(RRRp) + O(RRR)  $\to$   OLR2(RpR) 		k = 62.095095\\
429) OLR2(RpR)  $\to$   OL(RRRp) +  O(RRR) 		k = 0.008046\\
430) OL(RRRp) + O(CRR)  $\to$   OLR2(RpC) 		k = 62.095095\\
431) OLR2(RpC)  $\to$   OL(RRRp) +  O(CRR) 		k = 0.008046\\
432) OL(RRRp) + O(RpRR)  $\to$   OLR2(RpRp) 		k = 62.095095\\
433) OLR2(RpRp)  $\to$   OL(RRRp) +  O(RpRR) 		k = 0.008046\\
434) OLR2(00) + CI$_2$  $\to$   OLR2(R0) 		k = 0.314159\\
435) OLR2(R0)  $\to$   OLR2(00) +  CI$_2$ 		k = 0.035701\\
436) OLR2(00) + CI$_2$  $\to$   OLR2(0R) 		k = 0.314159\\
437) OLR2(0R)  $\to$   OLR2(00) +  CI$_2$ 		k = 38.256936\\
438) OLR2(00) + Cro$_2$  $\to$   OLR2(C0) 		k = 0.314159\\
439) OLR2(C0)  $\to$   OLR2(00) +  Cro$_2$ 		k = 0.662315\\
440) OLR2(00) + Cro$_2$  $\to$   OLR2(0C) 		k = 0.314159\\
441) OLR2(0C)  $\to$   OLR2(00) +  Cro$_2$ 		k = 0.068319\\
442) OLR2(00) + Rp  $\to$   OLR2(0Rp) 		k = 0.314159\\
443) OLR2(0Rp)  $\to$   OLR2(00) +  Rp 		k = 1.490712\\
444) OLR2(00) + Rp  $\to$   OLR2(Rp0) 		k = 0.314159\\
445) OLR2(Rp0)  $\to$   OLR2(00) +  Rp 		k = 2.062175\\
446) OLR2(R0) + CI$_2$  $\to$   OLR2(RR) 		k = 0.314159\\
447) OLR2(RR)  $\to$   OLR2(R0) +  CI$_2$ 		k = 38.256936\\
448) OLR2(R0) + Cro$_2$  $\to$   OLR2(RC) 		k = 0.314159\\
449) OLR2(RC)  $\to$   OLR2(R0) +  Cro$_2$ 		k = 0.068319\\
450) OLR2(R0) + Rp  $\to$   OLR2(RRp) 		k = 0.314159\\
451) OLR2(RRp)  $\to$   OLR2(R0) +  Rp 		k = 1.490712\\
452) OLR2(C0) + CI$_2$  $\to$   OLR2(CR) 		k = 0.314159\\
453) OLR2(CR)  $\to$   OLR2(C0) +  CI$_2$ 		k = 38.256936\\
454) OLR2(C0) + Cro$_2$  $\to$   OLR2(CC) 		k = 0.314159\\
455) OLR2(CC)  $\to$   OLR2(C0) +  Cro$_2$ 		k = 0.068319\\
456) OLR2(C0) + Rp  $\to$   OLR2(CRp) 		k = 0.314159\\
457) OLR2(CRp)  $\to$   OLR2(C0) +  Rp 		k = 1.490712\\
458) OLR2(0R) + CI$_2$  $\to$   OLR2(RR) 		k = 0.314159\\
459) OLR2(RR)  $\to$   OLR2(0R) +  CI$_2$ 		k = 0.035701\\
460) OLR2(0R) + Cro$_2$  $\to$   OLR2(CR) 		k = 0.314159\\
461) OLR2(CR)  $\to$   OLR2(0R) +  Cro$_2$ 		k = 0.662315\\
462) OLR2(0R) + Rp  $\to$   OLR2(RpR) 		k = 0.314159\\
463) OLR2(RpR)  $\to$   OLR2(0R) +  Rp 		k = 2.062175\\
464) OLR2(0C) + CI$_2$  $\to$   OLR2(RC) 		k = 0.314159\\
465) OLR2(RC)  $\to$   OLR2(0C) +  CI$_2$ 		k = 0.035701\\
466) OLR2(0C) + Cro$_2$  $\to$   OLR2(CC) 		k = 0.314159\\
467) OLR2(CC)  $\to$   OLR2(0C) +  Cro$_2$ 		k = 0.662315\\
468) OLR2(0C) + Rp  $\to$   OLR2(RpC) 		k = 0.314159\\
469) OLR2(RpC)  $\to$   OLR2(0C) +  Rp 		k = 2.062175\\
470) OLR2(0Rp) + CI$_2$  $\to$   OLR2(RRp) 		k = 0.314159\\
471) OLR2(RRp)  $\to$   OLR2(0Rp) +  CI$_2$ 		k = 0.035701\\
472) OLR2(0Rp) + Cro$_2$  $\to$   OLR2(CRp) 		k = 0.314159\\
473) OLR2(CRp)  $\to$   OLR2(0Rp) +  Cro$_2$ 		k = 0.662315\\
474) OLR2(0Rp) + Rp  $\to$   OLR2(RpRp) 		k = 0.314159\\
475) OLR2(RpRp)  $\to$   OLR2(0Rp) +  Rp 		k = 2.062175\\
476) OLR2(0Rp)  $\to$   OLR2(00) +  Rp +  MCI 		k = 0.011000\\
477) OLR2(RRp)  $\to$   OLR2(R0) +  Rp +  MCI 		k = 0.011000\\
478) OLR2(CRp)  $\to$   OLR2(C0) +  Rp +  MCI 		k = 0.011000\\
479) OLR2(RpRp)  $\to$   OLR2(Rp0) +  Rp +  MCI 		k = 0.011000\\
480) OL(0RR) + O(RR0  $\to$   OLR3(00) 		k = 62.095095\\
481) OLR3(00)  $\to$   OL(0RR) +  O(RR0) 		k = 0.008046\\
482) OL(0RR + O(RRR  $\to$   OLR3(0R) 		k = 62.095095\\
483) OLR3(0R)  $\to$   OL(0RR) +  O(RRR) 		k = 0.008046\\
484) OL(0RR) + O(RRC  $\to$   OLR3(0C) 		k = 62.095095\\
485) OLR3(0C)  $\to$   OL(0RR) +  O(RRC) 		k = 0.008046\\
486) OL(RRR) + O(RR0  $\to$   OLR3(R0) 		k = 62.095095\\
487) OLR3(R0)  $\to$   OL(RRR) +  O(RR0) 		k = 0.008046\\
488) OL(RRR) + O(RRR  $\to$   OLR3(RR) 		k = 62.095095\\
489) OLR3(RR)  $\to$   OL(RRR) +  O(RRR) 		k = 0.008046\\
490) OL(RRR) + O(RRC  $\to$   OLR3(RC) 		k = 62.095095\\
491) OLR3(RC)  $\to$   OL(RRR) +  O(RRC) 		k = 0.008046\\
492) OL(CRR) + O(RR0  $\to$   OLR3(C0) 		k = 62.095095\\
493) OLR3(C0)  $\to$   OL(CRR) +  O(RR0) 		k = 0.008046\\
494) OL(CRR) + O(RRR  $\to$   OLR3(CR) 		k = 62.095095\\
495) OLR3(CR)  $\to$   OL(CRR) +  O(RRR) 		k = 0.008046\\
496) OL(CRR) + O(RRC  $\to$   OLR3(CC) 		k = 62.095095\\
497) OLR3(CC)  $\to$   OL(CRR) +  O(RRC) 		k = 0.008046\\
498) OLR3(00) + CI$_2$  $\to$   OLR3(R0) 		k = 0.314159\\
499) OLR3(R0)  $\to$   OLR3(00) +  CI$_2$ 		k = 0.346100\\
500) OLR3(00) + CI$_2$  $\to$   OLR3(0R) 		k = 0.314159\\
501) OLR3(0R)  $\to$   OLR3(00) +  CI$_2$ 		k = 0.294263\\
502) OLR3(00) + Cro$_2$  $\to$   OLR3(C0) 		k = 0.314159\\
503) OLR3(C0)  $\to$   OLR3(00) +  Cro$_2$ 		k = 0.068319\\
504) OLR3(00) + Cro$_2$  $\to$   OLR3(0C) 		k = 0.314159\\
505) OLR3(0C)  $\to$   OLR3(00) +  Cro$_2$ 		k = 0.662315\\
506) OLR3(R0) + CI$_2$  $\to$   OLR3(RR) 		k = 0.314159\\
507) OLR3(RR)  $\to$   OLR3(R0) +  CI$_2$ 		k = 0.294263\\
508) OLR3(R0) + Cro$_2$  $\to$   OLR3(RC) 		k = 0.314159\\
509) OLR3(RC)  $\to$   OLR3(R0) +  Cro$_2$ 		k = 0.662315\\
510) OLR3(C0) + CI$_2$  $\to$   OLR3(CR) 		k = 0.314159\\
511) OLR3(CR)  $\to$   OLR3(C0) +  CI$_2$ 		k = 0.294263\\
512) OLR3(C0) + Cro$_2$  $\to$   OLR3(CC) 		k = 0.314159\\
513) OLR3(CC)  $\to$   OLR3(C0) +  Cro$_2$ 		k = 0.662315\\
514) OLR3(0R) + CI$_2$  $\to$   OLR3(RR) 		k = 0.314159\\
515) OLR3(RR)  $\to$   OLR3(0R) +  CI$_2$ 		k = 0.346100\\
516) OLR3(0R) + Cro$_2$  $\to$   OLR3(CR) 		k = 0.314159\\
517) OLR3(CR)  $\to$   OLR3(0R) +  Cro$_2$ 		k = 0.068319\\
518) OLR3(0C) + CI$_2$  $\to$   OLR3(RC) 		k = 0.314159\\
519) OLR3(RC)  $\to$   OLR3(0C) +  CI$_2$ 		k = 0.346100\\
520) OLR3(0C) + Cro$_2$  $\to$   OLR3(CC) 		k = 0.314159\\
521) OLR3(CC)  $\to$   OLR3(0C) +  Cro$_2$ 		k = 0.068319\\
522) OL(RR0) + O(RR0  $\to$   OLR4(00) 		k = 62.095095\\
523) OLR4(00)  $\to$   OL(RR0) +  O(RR0) 		k = 0.008046\\
524) OL(RR0) + O(RRR  $\to$   OLR4(0R) 		k = 62.095095\\
525) OLR4(0R)  $\to$   OL(RR0) +  O(RRR) 		k = 0.008046\\
526) OL(RR0) + O(RRC  $\to$   OLR4(0C) 		k = 62.095095\\
527) OLR4(0C)  $\to$   OL(RR0) +  O(RRC) 		k = 0.008046\\
528) OL(RRR) + O(RR0  $\to$   OLR4(R0) 		k = 62.095095\\
529) OLR4(R0)  $\to$   OL(RRR) +  O(RR0) 		k = 0.008046\\
530) OL(RRR) + O(RRR  $\to$   OLR4(RR) 		k = 62.095095\\
531) OLR4(RR)  $\to$   OL(RRR) +  O(RRR) 		k = 0.008046\\
532) OL(RRR) + O(RRC  $\to$   OLR4(RC) 		k = 62.095095\\
533) OLR4(RC)  $\to$   OL(RRR) +  O(RRC) 		k = 0.008046\\
534) OL(RRC) + O(RR0  $\to$   OLR4(C0) 		k = 62.095095\\
535) OLR4(C0)  $\to$   OL(RRC) +  O(RR0) 		k = 0.008046\\
536) OL(RRC) + O(RRR  $\to$   OLR4(CR) 		k = 62.095095\\
537) OLR4(CR)  $\to$   OL(RRC) +  O(RRR) 		k = 0.008046\\
538) OL(RRC) + O(RRC  $\to$   OLR4(CC) 		k = 62.095095\\
539) OLR4(CC)  $\to$   OL(RRC) +  O(RRC) 		k = 0.008046\\
540) OL(RRRp) + O(RR0  $\to$   OLR4(Rp0) 		k = 62.095095\\
541) OLR4(Rp0)  $\to$   OL(RRRp) +  O(RR0) 		k = 0.008046\\
542) OL(RRRp) + O(RRR  $\to$   OLR4(RpR) 		k = 62.095095\\
543) OLR4(RpR)  $\to$   OL(RRRp) +  O(RRR) 		k = 0.008046\\
544) OL(RRRp) + O(RRC  $\to$   OLR4(RpC) 		k = 62.095095\\
545) OLR4(RpC  $\to$   OL(RRRp) +  O(RRC) 		k = 0.008046\\
546) OLR4(00) + CI$_2$  $\to$   OLR4(R0) 		k = 0.314159\\
547) OLR4(R0)  $\to$   OLR4(00) +  CI$_2$ 		k = 0.035701\\
548) OLR4(00) + CI$_2$  $\to$   OLR4(0R) 		k = 0.314159\\
549) OLR4(0R)  $\to$   OLR4(00) +  CI$_2$ 		k = 0.294263\\
550) OLR4(00) + Cro$_2$  $\to$   OLR4(C0) 		k = 0.314159\\
551) OLR4(C0) $\to$   OLR4(00) +  Cro$_2$ 		k = 0.662315\\
552) OLR4(00) + Cro$_2$  $\to$   OLR4(0C) 		k = 0.314159\\
553) OLR4(0C)  $\to$   OLR4(00) +  Cro$_2$ 		k = 0.662315\\
554) OLR4(00) + Rp  $\to$   OLR4(Rp0) 		k = 0.314159\\
555) OLR4(Rp0)  $\to$   OLR4(00) +  Rp 		k = 2.062175\\
556) OLR4(R0) + CI$_2$  $\to$   OLR4(RR) 		k = 0.314159\\
557) OLR4(RR)  $\to$   OLR4(R0) +  CI$_2$ 		k = 0.002263\\
558) OLR4(R0) + Cro$_2$  $\to$   OLR4(RC) 		k = 0.314159\\
559) OLR4(RC)  $\to$   OLR4(R0) +  Cro$_2$ 		k = 0.662315\\
560) OLR4(C0) + CI$_2$  $\to$   OLR4(CR) 		k = 0.314159\\
561) OLR4(CR)  $\to$   OLR4(C0) +  CI$_2$ 		k = 0.294263\\
562) OLR4(C0) + Cro$_2$  $\to$   OLR4(CC) 		k = 0.314159\\
563) OLR4(CC)  $\to$   OLR4(C0) +  Cro$_2$ 		k = 0.662315\\
564) OLR4(0R) + CI$_2$  $\to$   OLR4(RR) 		k = 0.314159\\
565) OLR4(RR)  $\to$   OLR4(0R) +  CI$_2$ 		k = 0.000275\\
566) OLR4(0R) + Cro$_2$  $\to$   OLR4(CR) 		k = 0.314159\\
567) OLR4(CR)  $\to$   OLR4(0R) +  Cro$_2$ 		k = 0.662315\\
568) OLR4(0R) + Rp  $\to$   OLR4(RpR) 		k = 0.314159\\
569) OLR4(RpR)  $\to$   OLR4(0R) +  Rp 		k = 2.062175\\
570) OLR4(0C) + CI$_2$  $\to$   OLR4(RC) 		k = 0.314159\\
571) OLR4(RC)  $\to$   OLR4(0C) +  CI$_2$ 		k = 0.035701\\
572) OLR4(0C) + Cro$_2$  $\to$   OLR4(CC) 		k = 0.314159\\
573) OLR4(0C)  $\to$   OLR4(CC) +  Cro$_2$ 		k = 0.662315\\
574) OLR4(0C) + Rp  $\to$   OLR4(RpC) 		k = 0.314159\\
575) OLR4(RpC)  $\to$   OLR4(0C) +  Rp 		k = 2.062175\\
Including non-specific binding extends the reaction to scheme with these four reactions (rate constants calculated for $\Delta G_{\rm NSB}=-3.5$ kcal/mol):\\
576) D + CI$_2$  $\to$   1 DCI 		k = 0.314159\\
577) DCI  $\to$   1 D +  1 CI$_2$ 		k = 646634.937870\\
578) D + Cro$_2$  $\to$   1 DCRO 		k = 0.314159\\
579) DCRO  $\to$   1 D +  1 Cro$_2$ 		k = 646634.937870\\

\bibliographystyle{pnas}

\end{document}